\documentclass[a4paper,11pt]{article}
\pdfoutput=1
\usepackage{import}
\usepackage{jheppub}
\usepackage{hyperref}
\usepackage{natbib}
\usepackage{longtable}

\usepackage{amsmath,bm,amssymb,amsthm,mathrsfs,mathtools,slashed}
\usepackage{subcaption,color}
\usepackage{ragged2e}
\usepackage{caption}
\justifying\let\raggedright\justifying

\allowdisplaybreaks[4]
\usepackage{tikz}
\usepackage{multirow}
\usepackage{comment}
\usetikzlibrary{backgrounds}
\usetikzlibrary{shapes,matrix,trees}
\usetikzlibrary{arrows.meta}
\usepackage{centernot}
\usetikzlibrary{positioning}			% For "above of=" commands
\usetikzlibrary{calc,through}			% For coordinates
\usetikzlibrary{decorations.pathreplacing}     % For curly braces
\usepackage{pgffor}                                        % For repeating patterns
\usetikzlibrary{decorations.pathmorphing}	% For Feynman Diagrams
\usetikzlibrary{decorations.markings}
\tikzset{
	vector/.style={decorate, decoration={snake}, draw},
	provector/.style={decorate, decoration={snake,amplitude=2.5pt}, draw},
	antivector/.style={decorate, decoration={snake,amplitude=-2.5pt}, draw},
	fermion/.style={draw=black, postaction={decorate},
		decoration={markings,mark=at position .55 with {\arrow[draw=black]{>}}}},
	fermionbar/.style={draw=black, postaction={decorate},
		decoration={markings,mark=at position .55 with {\arrow[draw=black]{<}}}},
	fermionnoarrow/.style={draw=black},
	gluon/.style={decorate, draw=black,
		decoration={coil,amplitude=4pt, segment length=5pt}},
	scalar/.style={dashed,draw=black, postaction={decorate},
		decoration={markings,mark=at position .55 with {\arrow[draw=black]{>}}}},
	scalarbar/.style={dashed,draw=black, postaction={decorate},
		decoration={markings,mark=at position .55 with {\arrow[draw=black]{<}}}},
	scalarnoarrow/.style={dashed,draw=black},
	electron/.style={draw=black, postaction={decorate},
		decoration={markings,mark=at position .55 with {\arrow[draw=black]{>}}}},
	bigvector/.style={decorate, decoration={snake,amplitude=4pt}, draw},
	photon/.style={decorate, draw=black,decoration={snake,amplitude=4pt, segment length=5pt} }
}

\definecolor{ccblue}{rgb}{0.0,0.4,0.8}
\usepackage[]{hyperref}
\hypersetup{  colorlinks=true,
	linkcolor=ccblue,
	urlcolor=ccblue,
	citecolor=ccblue}
\usepackage{amsmath}
\usepackage{amsfonts}
\usepackage{amssymb}
\usepackage{booktabs}
\usepackage{slashed}
\usepackage[]{hyperref}
\usepackage{autobreak}
\newcommand{\nn}{\nonumber}
\newcommand{\ep}{\epsilon}
\newcommand{\lih}{{\rm Li}_4({\scriptscriptstyle \frac{1}{2}})}

\newcommand{\iomx}{\bar{x}}
\newcommand{\fxmo}{f_{\scriptscriptstyle -1}~}
\newcommand{\fxmosq}{f_{\scriptscriptstyle -1}^{2}~}
\newcommand{\fxt}{f_{\scriptscriptstyle 2}~}
\newcommand{\fxthr}{f_{\scriptscriptstyle 3}~}
\newcommand{\fxei}{f_{\scriptscriptstyle 8}~}
\newcommand{\fxtnt}{f_{\scriptscriptstyle 2,3}~}

\newcommand{\ixthr}{x^{{\scriptscriptstyle -}3}}
\newcommand{\ixtw}{x^{{\scriptscriptstyle -}2}}
\newcommand{\ixo}{x^{{\scriptscriptstyle -}1}}

\newcommand{\asr}{\Big( \frac{\alpha_s}{4\pi} \Big)}
\newcommand{\asrL}{\Big( \frac{\bar{\alpha}_s}{4\pi} \Big)}
\newcommand{\D}{{\cal D}}
\title{Three loop QCD corrections to the heavy-light form factors in the color-planar limit}

\author[a]{Sudeepan Datta,}
\author[b,c]{Narayan Rana,}
\author[d]{V. Ravindran,}
\author[d]{Ratan Sarkar}

\affiliation[a]{Centre for High Energy Physics, Indian Institute of Science, Bangalore 560012, India}
\affiliation[b]{School of Physical Sciences, National Institute of Science Education and Research,\\ An OCC of Homi Bhabha National Institute, Jatni 752050, India}
\affiliation[c]{Department of Physics, Indian Institute of Technology Kanpur, Kanpur 208016, India\footnote{On leave.}}
\affiliation[d]{The Institute of Mathematical Sciences, A CI of Homi Bhabha National Institute,\\ Taramani, Chennai 600113, India}

\emailAdd{sudeepand@iisc.ac.in, narayan.rana@niser.ac.in,\\
ravindra@imsc.res.in, 
ratansarkar@imsc.res.in}

\abstract{We present the analytic expressions for the color-planar contributions to the heavy-light form factors at three loops in perturbative QCD. 
These form factors play an important role in the precision predictions of various observables in top quark and flavour physics.
We compute the master integrals using the method of differential equations. We perform the ultraviolet renormalization for all the appearing fields and parameters. The analytic results for the renormalized form factors are expressed in terms of generalized harmonic polylogarithms.
We also study the Sudakov behaviour of these form factors in the asymptotic limit, which enables us to obtain the complete logarithmic three-loop and
partial four-loop contributions.
}

\newpage

\begin{document} 
\maketitle

\section{Introduction} 
Form factors encode the kinematical behaviour of scattering amplitudes in quantum field theories, and as such could be considered as the basic building blocks of the amplitudes.
Form factors not only provide important insights about the analytic structure of the scattering amplitudes,
but also play a central role in obtaining precise theoretical predictions for the physical observables
by contributing to their virtual corrections, and thus, in enhancing our ability to probe for new physics at large.

The form factors for various local composite operators between massless physical states (such as a pair of gluons or quarks), have been computed up to the fourth order in perturbative quantum chromodynamics (QCD) \cite{Ravindran:2004mb,deFlorian:2013sza,Moch:2005tm,Moch:2005id,Baikov:2009bg,Gehrmann:2010ue,Gehrmann:2014vha,
Ahmed:2015qia,Ahmed:2015qpa,Ahmed:2016vgl,Ahmed:2016qjf,Ahmed:2019yjt,Lee:2021lkc,Lee:2022nhh,Chakraborty:2022yan}, and have played a key role in precision studies of the Standard Model (SM) parameters.
The heavy quark form factors (HQFF) are the cases when the physical states are massive (such as a pair of top quarks),
and are of critical importance in the study of physical observables - such as the forward-backward asymmetry of heavy quark pair production at lepton colliders, the study of anomalous magnetic moment of a heavy quark, and more importantly, the study of top-quark properties at the High-Luminosity Large Hadron Collider (HL-LHC). The current state-of-the-art computations in this frontier involve third order QCD $\big(\mathcal{O}(\alpha_s^3)\big)$ corrections \cite{Bernreuther:2004ih,Bernreuther:2004th,Bernreuther:2005rw,Bernreuther:2005gw,Gluza:2009yy,Ablinger:2017hst,Henn:2016tyf,Lee:2018nxa,Ablinger:2018yae,Lee:2018rgs,Ablinger:2018zwz,Blumlein:2019oas,Fael:2022rgm,Fael:2022miw,Fael:2023zqr,Blumlein:2023uuq} to the HQFF.

The heavy-light form factors (HLFF) play an important role in analysing the properties of heavy quarks, particularly the flavour-changing decays of such quarks. With appropriate analytic continuation, these form factors constitute the virtual corrections for a heavy quark decaying into a light quark and a charged ($W$) boson. Few examples of such decays are $t \rightarrow b W^{*}$, $b \rightarrow c W^{*}$ and $b \rightarrow u W^{*}$.
The virtual corrections to the $t \rightarrow b W^{*}$ decay are very important 
to obtain a precise top-quark decay width $\Gamma_t$, one of the most important parameters in the studies of fundamental interactions. This is particularly relevant for top-quark studies to obtain an understanding of its scope in opening up a portal to beyond the SM (BSM) physics, and as such, both the production and decay modes of the top are still subjected to active theoretical and experimental scrutiny, particularly from a precision point of view. The other two processes contribute to the hard functions that are required for calculations in the Soft Collinear Effective Theory (SCET), to evaluate the differential distribution of the inclusive semi-leptonic $B$ meson decays in the shape-function region. This allows a precise determination of the Cabibbo–Kobayashi–Maskawa (CKM) matrix elements $|V_{cb}|$ and $|V_{ub}|$, which are relevant for the study of flavour and CP-violation in the quark sector.
The top-quark decay, due to its high importance, has already been studied up to next-to-next-to leading order (NNLO) in QCD \cite{Chetyrkin:1999ju,Blokland:2004ye,Gao:2012ja,Brucherseifer:2013iv}. 
Within the regime of SCET, the semi-leptonic $B$ decays have been studied up to NNLO QCD as well \cite{Bosch:2004th,Lange:2005yw,Becher:2005pd,Becher:2006qw,Asatrian:2008uk,Beneke:2008ei,Bell:2008ws}. 
The explicit two-loop QCD corrections to the HLFF have been presented in refs.~\cite{Bonciani:2008wf,Huber:2009se,Bell:2006tz}.  
To keep up with the increasing volume of precision measurements at the collider experiments, it is essential to improve the precision predictions from the theory side. At the same time, it is imperative that a uniform precision is achieved for both the production and decay processes, which in turn necessitates the inclusion of third order QCD corrections.
To this end, the three-loop master integrals (MIs) relevant to obtain the color-planar limit of the HLFF have been computed in \cite{Chen:2018fwb}.

In this work, we provide for the first time, the analytic results for three-loop QCD corrections to the HLFF in the color-planar limit. We use the method of 
integration-by-parts (IBP) reduction technique \cite{Chetyrkin:1981qh,Laporta:2001dd} to obtain a basis of MIs for the process under consideration. These MIs are analytically evaluated through the method of differential equations \cite{Kotikov:1990kg,Argeri:2007up,Remiddi:1997ny,Henn:2013pwa,Ablinger:2015tua}. 
We perform the ultraviolet (UV) renormalization in a mixed scheme, and eventually show that the UV renormalized form factors satisfy the universal infrared (IR) behaviour \cite{Becher:2009kw} - which provides an important consistency check of the UV renormalized results.
In addition, we also perform a rigorous study on the Sudakov behaviour of the HLFF in the asymptotic limit. Due to the universality of IR behaviour
and factorizations of the QCD scattering amplitudes, we can predict the high-energy logarithms by studying the integro-differential Sudakov equation
and the renormalization group equations (RGE). The study of Sudakov behaviour has been performed in case of 
massless form factors in refs.~\cite{Sudakov:1954sw,Mueller:1979ih,Collins:1980ih,Sen:1981sd,Ravindran:2005vv,Ravindran:2006cg,Ajjath:2019vmf}
and of HQFF in refs.~\cite{Mitov:2006xs,Ahmed:2017gyt,Blumlein:2018tmz}.
We present the first such study for the HLFF in this paper.

This paper is organized as follows. In section \ref{sec:amps}, we discuss the theoretical  basics of the process under consideration i.e. the definition of the form factors, the relevant projectors, the procedure of UV renormalization and the universal IR structure. In section \ref{sec:comp}, we briefly describe the details of the computational procedure, in particular, the exact computation of the MIs. In section \ref{sec:results}, we present the primary result
of this paper i.e. the three-loop color-planar finite remainders after performing appropriate IR subtraction. Next, we report 
a concise theoretical formulation for the study of Sudakov behaviour along with the three-loop results in section \ref{sec:asymp}.
Finally we conclude in section \ref{sec:conclusions}.
In the appendices \ref{sec:res1L}-\ref{sec:res3L}, we present the one, two and three-loop UV renormalized form factors. 
In appendix \ref{sec:res4L}, we also present partial four-loop results in the asymptotic limit.

\section{The heavy-light form factors}
\label{sec:amps}
We consider the decay process 
\begin{equation}
    t (P) \rightarrow  b (p) + W^* (q) \,.
\end{equation}
The top quark, carrying momentum $P$, decays into a massless bottom quark
and a off-shell $W$ boson. The bottom quark carries a momentum $p$ and the $W$ boson 
carries a momentum $q = P-p$. The on-shell conditions are given by
\begin{equation}
    P^2 = m_t^2\,, ~~ p^2 = 0 \,.
\end{equation}
$m_t$ denotes the mass of the top quark. 
We also define the dimensionless variable $x$ as
\begin{equation}
    x = \frac{q^2}{m_t^2} \,.
\end{equation}
The general form of the amplitude is given by
\begin{equation}
 \bar{b}_c (p) ~ \Gamma^{\mu}_{cd} ~ t_d (P)
\end{equation}
where $\bar{b}_c (p)$ and $t_d (P)$ are the bi-spinors of the $b$-quark and $t$-quark, respectively.
The generic structure of the vertex with one massive and one massless quark
can be described in terms of three independent form factors\footnote{The definitions of our $G_2$ and $G_3$ have been swapped, as compared to the ones presented in \cite{Bonciani:2008wf}.} 
as
\begin{equation}
 \Gamma^{\mu}_{cd} = -i \delta_{cd} \Big[ G_1 \gamma^\mu (1-\gamma_5) + \frac{G_2}{2 m_t} (1+\gamma_5) ( P^\mu + p^\mu )
    + \frac{G_3}{2 m_t} (1+\gamma_5) ( P^\mu - p^\mu ) \Big] \,.
\end{equation}
Each form factor ($G_i$) can be obtained by applying appropriate projector
on the amplitudes and performing trace over the spinor and color indices. 
We find the projectors (${\cal{P}}_i$) to be applied on $\Gamma^{\mu}_{cd}$ for the corresponding form factor ($G_i$) as
\begin{equation}
  {\cal{P}}_i = -\frac{i}{N_C} \delta_{cd} (\slashed{P}+m_t) \Big[ g_{i,1}  \gamma^\mu (1-\gamma_5) - \frac{g_{i,2}}{2 m_t} (1-\gamma_5) ( P^\mu + p^\mu )
     - \frac{g_{i,3}}{2 m_t} (1-\gamma_5) ( P^\mu - p^\mu ) \Big] \slashed{p}
\end{equation}
where
\begin{align}
    g_{1,1} &= \frac{1}{4 m_t^2 (d-2) (1-x)}  \,,
    g_{1,2} = \frac{1}{2 m_t^2 (d-2) (1-x)^2}  \,,
    g_{1,3} = -\frac{1}{2 m_t^2 (d-2) (1-x)^2}  \,,
    \nn \\
    g_{2,1} &= -\frac{1}{2 m_t^2 (d-2) (1-x)^2}  \,,
    g_{2,2} = -\frac{1+(d-2)x}{m_t^2 (d-2) (1-x)^3} \,,
    g_{2,3} = \frac{d-1}{m_t^2 (d-2) (1-x)^3}  \,,
    \nn \\
    g_{3,1} &= \frac{1}{2 m_t^2 (d-2) (1-x)^2} \,,
    g_{3,2} = \frac{d-1}{m_t^2 (d-2) (1-x)^3} \,,
    g_{3,3} = -\frac{(2 d-3)-(d-2)x}{m_t^2 (d-2) (1-x)^3} \,.
\end{align}
$N_C$ denotes the number of colors. We also mention here the Casimir operators $C_A = N_C$, $C_F = \frac{N_C^2-1}{2N_C}$ and $T_F = \frac{1}{2}$ for SU($N_C$).
We denote the number of light and heavy quarks by $n_l$ and $n_h$, respectively.
The form factors ($G_i$) can be expanded in powers of the strong coupling constant ($\alpha_s$) as 
\begin{equation}
    G_i = \frac{i g_w}{2 \sqrt{2}} \sum_{n=0}^\infty \Big( \frac{\alpha_s}{4\pi} \Big)^n G_i^{(n)}
\end{equation}
where $g_w = \frac{e}{s_w}$ is the weak interaction coupling constant.
$e$ is the electric charge and $s_w$ is the sine of the weak mixing angle.
Accordingly the leading order form factors are
\begin{equation}
    G_1^{(0)} = 1, \quad G_2^{(0)} = 0, \quad G_3^{(0)} = 0.
\end{equation}

\subsection{Ultraviolet renormalization}
We use dimensional regularization in $d=4-2\epsilon$ space-time dimensions
to regularize the bare form factors.
As the vertex is chiral, it becomes necessary to establish the treatment of $\gamma_5$ within this regularization scheme. 
However, both the $\gamma_5$ from the vertex and the projectors, are connected to the open fermion lines.
Hence, we use the prescription presented in \cite{Kreimer:1989ke,Korner:1991sx} i.e. the anti-commutation of $\gamma_5$ with other $\gamma$-matrices (\{$\gamma_\mu,\gamma_5\} = 0$),
followed by $\gamma_5^2 = 1$.

We follow a mixed scheme to perform the UV renormalization of the form factors. 
The mass and wave function of the heavy quark ($t$), and the wave function of the light quark ($b$)
have been renormalized in the on-shell (OS) renormalization scheme.
On the other hand, the strong coupling constant has been renormalized in the modified minimal subtraction ($\overline{\text{MS}}$) scheme \cite{tHooft:1973mfk,Bardeen:1978yd}.
We denote the corresponding renormalization constants by $Z_m, Z_{2,t}, Z_{2,b}$ and $Z_{\alpha_s}$, respectively.
All the renormalization constants follow a series expansion in $\alpha_s$ as 
\begin{equation}
 Z_i = \sum_{n=0}^\infty \Big( \frac{\alpha_s}{4 \pi} \Big)^n Z_i^{(n)} \,. 
\end{equation}
The required three-loop results are available for $Z_m$ in refs.~\cite{Broadhurst:1991fy,Melnikov:2000zc,Marquard:2007uj,Marquard:2015qpa,Marquard:2016dcn},
for $Z_{2,t}$ in refs.~\cite{Broadhurst:1991fy, Melnikov:2000zc,Marquard:2007uj,Marquard:2018rwx} 
and for $Z_{\alpha_s}$  in refs.~\cite{Tarasov:1980au,Larin:1993tp,vanRitbergen:1997va,Czakon:2004bu,Baikov:2016tgj,Herzog:2017ohr,Luthe:2017ttg}.
The $Z_{2,b}$ was presented in ref.~\cite{Bonciani:2008wf} up to two-loop and ${\mathcal O}(\ep^0)$.
The closed-form solution of $Z_{2,b}^{(2)}$ in $\ep$ was given 
in refs.~\cite{Beneke:2008ei,Bell:2008ws}.
To obtain the two-loop results up to ${\mathcal O}(\ep^2)$, we also re-compute $Z_{2,b}^{(2)}$ and present up to ${\mathcal O}(\ep^2)$ in the following
\begin{equation}
 Z_{2,b}^{(2)} = C_F n_h T_F \bigg[ \frac{1}{\ep} - \frac{5}{6} 
 + \ep \left( \frac{89}{36} + \zeta_2 \right)
 + \ep^2 \left( -\frac{869}{216} - \frac{5}{6} \zeta_2 - \frac{2}{3} \zeta_3 \right) 
 + {\mathcal O}(\ep^3) \bigg] \,.
\end{equation}
The mass renormalization of the heavy quark requires explicit computation of the Feynman diagrams after counter-term insertion.
The other renormalizations are multiplicative i.e. the renormalized form factors ($G_i$)
are related to the bare form factors ($\hat{G}_i$) as follows
\begin{equation}
 G_i = Z_{2,t}^{\frac{1}{2}} Z_{2,b}^{\frac{1}{2}} ( \hat{G}_i + \hat{G}_{ct,i} )\,.
\end{equation}
$\hat{G}_{ct,i}$ denotes the appropriate mass renormalization counter-term contributions from 
lower orders in $\alpha_s$. We also note that the strong coupling constant renormalization 
needs to be performed considering full QCD i.e. the corresponding $\beta$-functions are 
given by
\begin{align}
 \beta_0 &= \frac{11}{3} C_A - \frac{4}{3} T_F (n_h + n_l) \,,
 \nonumber\\
 \beta_1 &= \frac{34}{3} C_A^2 - \frac{20}{3} C_A T_F (n_h+n_l) - 4 C_F T_F (n_h+n_l) \,.  
\end{align}

\subsection{Ward identity}

The UV renormalized form factors satisfy the on-shell Ward identity
\begin{equation}
 q_\mu \Gamma^\mu - m_W \Gamma_{\text{PS}} = 0 \,.
\end{equation}
The pseudo-scalar vertex appears for the process where a top quark decays into a bottom quark and 
a charged pseudo-Goldstone boson. The generic structure of this vertex can be described as follows
with the form factor $S$
\begin{equation}
 \Gamma_{\text{PS}} = \frac{m_t}{m_W} S (1+\gamma_5) \,.
\end{equation}
In order to satisfy the Ward identity, it is necessary to renormalize the pseudo-scalar vertex as well.
Apart from the usual quantities i.e. the mass and the wave function of the heavy quark, the wave 
function of the light quark and the strong coupling constant, an additional renormalization is needed 
for the heavy quark mass ($m_t$) appearing in the vertex. This overall mass renormalization has been performed also in OS scheme. 
The UV renormalized form factor $S$ also can be expanded 
in $\alpha_s$ as
\begin{equation}
 S = \frac{i g_w}{2 \sqrt{2}} \sum_{n=0}^\infty \Big( \frac{\alpha_s}{4\pi} \Big)^n S^{(n)} \,.
\end{equation}
In terms of the form factors, the Ward identity takes the following form
\begin{equation}
\label{eq:WI}
 2 G_1^{(n)} + G_2^{(n)} + x G_3^{(n)} - 2 S^{(n)} = 0 \,.
\end{equation}

\subsection{Universal infrared behaviour}
The UV renormalized form factors still contain divergences of IR origin, arising from soft gluons and collinear partons.
However, they follow a universal structure. In case of massless scattering amplitudes, a plethora of 
seminal works \cite{Catani:1998bh,Sterman:2002qn,Becher:2009cu,Gardi:2009qi,Ravindran:2006cg}
has already provided a good understanding on the universal IR structure generated through the soft and collinear dynamics. 
A first step was taken in ref.~\cite{Mitov:2006xs} to achieve the universal IR structure for the two-loop scattering amplitudes 
with massive partons in the high energy limit. Later, it has been studied also at the three-loop level in refs.~\cite{Ahmed:2017gyt,Blumlein:2018tmz}.
Following a SCET approach,
the universal IR structures for a generic two-loop amplitude with both massless and massive partons, have been presented in ref.~\cite{Becher:2009kw},
which can easily be generalized to three loops for the form factors.

Following ref.~\cite{Becher:2009kw}, the IR singularities of the form factors can be factorized as a multiplicative renormalization factor,
as follows
\begin{equation}
 G_i = Z(\bar{\mu}) G_i^{\text{fin}} (\bar{\mu}) \,.
 \label{eq:ir}
\end{equation}
$G_i^{\text{fin}} (\bar{\mu})$ is finite as $\ep \rightarrow 0$.
$\bar{\mu}$ is the scale introduced for this particular factorization of IR singularities.
The structure of $Z(\bar{\mu})$ is controlled by the RGE which
is characterized by the corresponding anomalous dimension. However, before we proceed further,
we note, as also mentioned in \cite{Blumlein:2019oas}, that the anomalous dimensions have been computed 
considering only the massless QCD corrections with $n_l$ light quark flavours.
On the other hand, the form factors are defined considering full-QCD i.e.
$n_l$ light and $n_h$ heavy quark flavours. Hence, to describe the IR structure 
of the form factors, we solve the RGE for $\bar{Z}$, the equivalent of $Z$ in the 
massless QCD with $n_l$ light quark only and then use the decoupling relation for
the strong coupling constant i.e. the relation between $\bar{\alpha}_s$ and $\alpha_s$,
which are defined in the massless and the full QCD, respectively. 
The RGE for $\bar{Z}$ is
\begin{equation}
 \frac{d}{d \ln \bar{\mu}} \ln \bar{Z} (\ep,q^2,m_t,\bar{\mu}) = - \Gamma (q^2,m_t,\bar{\mu}) \,.
\end{equation}
$\Gamma$ is the corresponding anomalous dimension. 
In case of heavy quark form factors, $\Gamma$ is purely the massive cusp 
anomalous dimension \cite{Korchemsky:1987wg,Korchemsky:1991zp,Grozin:2014hna,Grozin:2015kna}.
For the case of HLFF, $\Gamma$ is a combination of massive and massless cusp anomalous dimensions.
To solve the RGE, we expand
both $\bar{Z}$ and $\Gamma$ in a perturbative series in $\alpha_s$ as follows
\begin{equation}
 \bar{Z} = \sum_{n=0}^\infty \Big( \frac{\bar{\alpha}_s}{4\pi} \Big)^n \bar{Z}^{(n)}\,,
 \quad
 \Gamma = \sum_{n=0}^\infty \asrL^{n+1} \Gamma_n \,.
\end{equation}
The solution can then be found as 
\begin{align}
 \ln \bar{Z} &= \asrL \bigg[ \frac{\Gamma_0'}{4 \ep^2}  + \frac{\Gamma_0}{2 \ep} \bigg]
       + \asrL^2 \bigg[ - \frac{3 \bar{\beta}_0 \Gamma_0'}{16 \ep^3} + \frac{\Gamma_1' - 4 \bar{\beta}_0 \Gamma_0}{16 \ep^2} + \frac{\Gamma_1}{4 \ep} \bigg]
       \nn\\
       &+ \asrL^3 \bigg[ \frac{11 \bar{\beta}_0^2 \Gamma_0'}{72 \ep^4} - \frac{5 \bar{\beta}_0 \Gamma_1' 
       + 8 \bar{\beta}_1 \Gamma_0' - 12 \bar{\beta}_0^2 \Gamma_0}{72 \ep^3}
                      + \frac{\Gamma_2' - 6 \bar{\beta}_0 \Gamma_1 - 6 \bar{\beta}_1 \Gamma_0}{36 \ep^2} + \frac{\Gamma_2}{6 \ep} \bigg]
       \nn\\
       &+ {\cal O}(\alpha_s^4) \,,
\end{align}
where, 
\begin{equation}
 \Gamma_n' = \frac{\partial}{\partial \bar{\mu}} \Gamma_n \,.
\end{equation}
$\bar{\beta}$ is the $\beta$-function in the massless QCD with $n_l$ light quarks.
To obtain $Z$ from $\bar{Z}$, we use the decoupling relation obtained using background field method 
in refs.~\cite{Chetyrkin:1997un,Grozin:2007fh,Grozin:2011nk}.

\section{Computational details}
\label{sec:comp}
We follow the conventional procedure to compute the HLFF starting with the generation of the corresponding Feynman
diagrams using \textsc{QGRAF} \cite{Nogueira:1991ex}. 
Considering only the QCD corrections we obtain 13 diagrams at two-loop and 263 diagrams at three-loop. 
We note that we have considered only one additional light quark to generate the diagrams, which is 
sufficient to compute the diagrams for all flavours.
We use in-house \textsc{FORM} \cite{Tentyukov:2007mu} routines to translate the \textsc{QGRAF} output into Feynman amplitudes and for further manipulations
of Dirac, Lorentz and color algebra and also to project out the form factors using the projectors.
The color algebra has also been performed using \textsc{Color} \cite{vanRitbergen:1998pn}.
We also use another independent approach based on \textsc{FeynArts} \cite{Hahn:2000kx}, \textsc{FeynCalc} \cite{Shtabovenko:2020gxv} and \textsc{Mathematica}
to generate the Feynman diagrams and perform the algebraic manipulations to compute the form factors up to two-loop,
finding perfect agreements.

After these first manipulations, the expressions contain a large number of scalar Feynman integrals.
Following standard procedure, we reduce them to a much smaller set of scalar integrals, the MIs
using IBP \cite{Chetyrkin:1981qh,Laporta:2001dd} identities. 
We have performed the reduction with two independent in-house programs based on the
public codes \textsc{Kira} \cite{Klappert:2020nbg} and \textsc{LiteRed} \cite{Lee:2013mka}.
In the next section, we present the details of the reduction.

\subsection{Integral families and reduction to master integrals}
All the Feynman diagrams contributing to the three-loop color-planar corrections of the HLFF
can be mapped to a single integral family. 
We note that 13 sectors in this integral family completely span all the appearing scalar integrals.
To set our convention, we present a generic three-loop integral as follows:
\begin{equation}
    I_{\nu}(d,x)= \int \prod_{i=1}^3 \frac{d^d k_i}{ (2 \pi)^{d}} \prod_{j=1}^{12} \frac{1}{D_{j}^{\nu_j}}
\end{equation}
where, $\nu = \nu_1\nu_2...\nu_{12}$.
We put $S_\ep = \exp ( - \ep (\gamma_E - \ln (4\pi)))$ equals to one for each loop order.
We choose the following combination of the propagators as our color-planar integral family
\begin{equation}
 \{ \D_1-m_t^2, \D_2-m_t^2, \D_3-m_t^2, \D_{12}, \D_{23}, \D_{13}, \D_{1;1}, \D_{2;1}, \D_{3;1}, \D_{1;12}, \D_{2;12}, \D_{3;12} \} 
\end{equation}
where,
\begin{equation}
 \D_i = k_i^2\,, \D_{ij} = (k_i-k_j)^2\,, \D_{i;1} = (k_i-P)^2\,, \D_{i;12} = (k_i-P+p)^2\,.   
\end{equation}
Using \textsc{Kira}, we have obtained $70$ MIs for this integral family, as listed in Table~\ref{table:mis}, classified as per the number of propagators appearing in the denominator.
\begin{table}
\begin{minipage}{0.48\textwidth}
\begingroup
 \centering
 \begin{longtable}{ |p{0.15cm}|p{0.8cm}| p{0.65\textwidth}| }
  \hline 
   \# & sector & \hfil master integrals  \\  
   \hline 
   \hline 
 3  & 7  &   $I_{111000000000}$  \\   
 \hline
 \hline
 4  & 29 &   $I_{101110000000}$  \\
    & 78 &   $I_{011100100000}$   \\
    & 92 &   $I_{001110100000}$   \\
    & 519 &  $I_{111000000100}$   \\
    & 526 &  $I_{011100000100}$, $I_{(-1)11100000100}$  \\
    & 540 &  $I_{001110000100}$, $I_{(-1)01110000100}$  \\
 \hline  
 \hline
 5  & 110 &   $I_{011101100000}$   \\
    & 244 &   $I_{001011110000}$   \\
    & 247 &   $I_{111011110000}$   \\
    & 541 &   $I_{101110000100}$   \\
    & 558 &   $I_{011101000100}$, $I_{(-1)11101000100}$  \\
    & 653 &   $I_{101100010100}$   \\
    & 661 &   $I_{101010010100}$   \\
    & 668 &   $I_{001110010100}$   \\
    & 684 &   $I_{001101010100}$, $I_{(-1)01101010100}$   \\
    & 689 &   $I_{100011010100}$   \\
    & 692 &   $I_{001011010100}$, $I_{(-1)01011010100}$   \\
    & 1543 &   $I_{111000000110}$   \\
    & 1557 &   $I_{101010000110}$, $I_{1(-1)1010000110}$   \\
    & 1588 &   $I_{001011000110}$, $I_{(-1)01011000110}$   \\
 \hline
 \hline
8   &  &   \\
 \hline
 \hline
9   & 1918 &  $I_{011111101110}$, $I_{(-1)11111101110}$\\   
 \hline 
 \end{longtable}
\endgroup
\end{minipage}
\begin{minipage}{0.49\textwidth}
\begingroup
 \centering
 \begin{longtable}{ |p{0.15cm}|p{0.8cm}| p{0.7\textwidth}| }
  \hline 
   \# & sector & \hfil master integrals  \\  
   \hline 
   \hline
6   & 655 &  $I_{111100010100}$, $I_{111100(-1)10100}$\\
    & 669 &  $I_{101110010100}$, $I_{1(-1)1110010100}$,\\
    &     &  $I_{10111(-1)010100}$,\,$I_{101110(-1)10100}$,\\
    &     &  $I_{1011100101(-1)0}$\\ [0.4ex]
%     %&     &  $I_{10111(-1)010100}$, $I_{101110(-1)10100}$,\\
%     %&     &  $I_{1011100101(-1)0}$\\
    & 686 &  $I_{011101010100}$, $I_{(-1)11101010100}$,\\
    &     &  $I_{0111(-1)1010100}$, $I_{011101(-1)10100}$\\
%     %&     &  $I_{0111(-1)1010100}$, $I_{011101(-1)10100}$\\
    & 691 &  $I_{110011010100}$, $I_{11(-1)011010100}$\\
    & 693 &  $I_{101011010100}$, $I_{1(-1)1011010100}$\\
    & 694 &  $I_{011011010100}$, $I_{(-1)11011010100}$\\
    & 700 &  $I_{001111010100}$, $I_{(-1)01111010100}$\\
    & 937 &  $I_{100101011100}$\\
    & 1587 &  $I_{110011000110}$\\
    & 1811 &  $I_{110010001110}$\\
    & 1841 &  $I_{100011001110}$\\
    & 3591 &  $I_{111000000111}$\\ [0.5ex]
 \hline
 \hline
7   & 695 &  $I_{111011010100}$, $I_{111(-1)11010100}$,\\
    &     &  $I_{111011(-1)10100}$, $I_{1110110101(-1)0}$\\ [0.4ex]
% %   &     &  $I_{111011(-1)10100}$, $I_{1110110101(-1)0}$\\
    & 939 &  $I_{110101011100}$, $I_{11(-1)101011100}$\\
    & 1591 &  $I_{111011000110}$, $I_{111(-1)11000110}$\\
    & 1654 &  $I_{011011100110}$, $I_{011(-1)11100110}$\\
    & 1815 &  $I_{111010001110}$, $I_{11101(-1)001110}$\\
    & 1821 &  $I_{101110001110}$, $I_{10111(-1)001110}$\\
    & 1845 &  $I_{101011001110}$, $I_{1(-1)1011001110}$\\[0.5ex]
 \hline
 \end{longtable}
\endgroup
\end{minipage}
    \caption{List of the master integrals. \# indicates the number 
    of propagators.}
    \label{table:mis}
\end{table}

\subsection{Computation of the master integrals}
In this section, we summarize the computational details of the MIs. We follow the generic method of 
differential equations 
\cite{Kotikov:1990kg,Argeri:2007up,Remiddi:1997ny,Henn:2013pwa,Ablinger:2015tua}. 
To set up the system of differential equations of the MIs, we differentiate them with respect to the variable $x$,
followed by IBP reductions on the differentiated output to obtain a linear 
combination of MIs for each differentiated integral.
We obtain a $70 \times 70$ system of coupled linear differential equations, which can be symbolically presented as
% 
%-----------------------------------------------------------------------------------------------------------------------------
\begin{equation}\label{Equ:InputSystem}
 \frac{d}{dx} {\cal I} = {\cal M ~ I}.
\end{equation}
%-----------------------------------------------------------------------------------------------------------------------------
Here the $70 \times 70$ matrix ${\cal M}$ contains rational functions in $x$ and $d$.
To proceed, we first rearrange the system of differential equations in an upper block-triangular form. 
In such an arrangement, the integrals in the last block are homogeneously coupled. Hence, we
compute the MIs block-by-block starting from the last one.
Also while solving for each block, we perform series expansion 
of the whole subsystem in $\ep$ and solve it order by order in $\ep$ successively, starting with the highest pole in $\ep$. 
At each order in $\ep$, we decouple the coupled subsystem 
and obtain higher-order differential equation(s), which has(have) been solved using the method of variation of constant.
In our case, the spanning alphabet is
%-----------------------------------------------------------------------------------------------------------------------------
\begin{equation}
\label{eq:LETT}
\bigg\{
\frac{1}{x},~~
\frac{1}{1-x},~~
\frac{1}{1+x},~~
\frac{1}{2-x}  \bigg\}
% \frac{x}{1-x+x^2},
\end{equation}
%-----------------------------------------------------------------------------------------------------------------------------
i.e.~the usual harmonic polylogarithms (HPLs) \cite{Remiddi:1999ew} and 
generalized HPLs \cite{Ablinger:2013cf}. 
The kernel with letter $2$ is defined such that
\begin{equation}
 H_2 (x) = \ln (2) - \ln (2-x) \,.
\end{equation}
The integration constants are determined using boundary conditions either at $x=0$ or $x=1$, or demanding regularity at $x=0$. 
For several two- and three-loop integrals, the boundary conditions 
have been computed using \textsc{MBConicHulls} \cite{Ananthanarayan:2020fhl,Banik:2022bmk} with \textsc{AMBRE} \cite{Gluza:2007rt}, and \textsc{HypExp} \cite{Huber:2005yg,Huber:2009se}.
We have used \textsc{HarmonicSums} \cite{Ablinger:2010kw,Ablinger:2011te,Ablinger:2014rba}, 
and 
\textsc{PolyLogTools} \cite{Duhr:2019tlz} at several stages of the computations.
All the computed MIs have been checked numerically against the dedicated numerical programs \textsc{FIESTA} \cite{Smirnov:2008py, Smirnov:2021rhf} and \textsc{AMFlow} \cite{Liu:2022chg}.

\subsubsection{A comment on the boundary conditions}
As noted in ref.~\cite{Chen:2018fwb}, the regularity condition provides one boundary constant for the $2\times2$ subsystem
of the MIs with 9 propagators.
To solve this subsystem, the other boundary constant must be computed explicitly. Instead of using the conventional 
methods, we have used the method of auxiliary mass flow \cite{Liu:2017jxz,Liu:2021wks} through \textsc{AMFlow} to compute the boundary constants
of these MIs at $x=0$ with 100 digits precision. 
As the appearing constants are only multiple zeta values (MZVs) \cite{Blumlein:2009cf}, 
we have used 
the \textsc{PSLQ} algorithm \cite{pslq:92} to generate the analytic 
expressions for the boundary constants, which agree with the results presented in \cite{Chen:2018fwb}.
Below, we have listed the integrals used to evaluate the boundary constants:
\begin{align}
 & J_1 := I_{011111101110}\,, ~~ J_2 := I_{0111111(-1)111(-1)}\,, ~~ J_3 := I_{01111110111(-1)} \,.
\end{align}
We present below only 60 digits of their corresponding numerical values as obtained up to 100 digits using \textsc{AMFlow}. We also present the analytic expressions for those integrals in terms of MZVs as obtained using \textsc{PSLQ}:
\begin{align}
 J_1 =&~ 0.0291666666666666666666666666666666666666666666666666666666667/\ep^4
 \nn\\
 &- 0.272222222222222222222222222222222222222222222222222222222222/\ep^3
 \nn\\
 &+ 2.59825692198426295563240247140224666960871026564183564789818/\ep^2
 \nn\\
 &- 14.3991851789832694696508356174382924385264868134461722268087/\ep
 \nn\\
 &+ 102.228914656763448459622280957586653192203871348934296116490 
 \nn\\
 &- 557.393658369305674148828733349502449840866254759521243337465 ~\ep 
%%%%%%%%%%
 \nn \\
 =&~ \frac{1}{\ep^4} \bigg( \frac{7}{240} \bigg) 
 + \frac{1}{\ep^3} \bigg( -\frac{49}{180} \bigg) 
 + \frac{1}{\ep^2} \bigg( \frac{1309}{720} +\frac{683 \zeta_2}{1440} \bigg) 
 \nn\\
 &+ \frac{1}{\ep} \bigg( -\frac{1997}{180} -\frac{649}{180} \zeta_2 +\frac{1573}{720} \zeta_3 \bigg) 
 + \bigg( \frac{9805}{144}+\frac{3411}{160} \zeta_2
     +\frac{199333 \zeta_2^2}{28800}-\frac{5887}{360} \zeta_3 \bigg) 
 \nn\\    
 &+ \ep \bigg( 
 -\frac{39151}{90}
-\frac{1211}{10} \zeta_2
-\frac{80939 \zeta_2^2}{1440}
+\frac{71737}{720} \zeta_3
+\frac{20587 \zeta_2 \zeta_3}{1440}
+\frac{93589 \zeta_5}{1200}
 \bigg)    
     \,.
% \end{align}
%
\\
%
% \begin{align}
 J_2 =& ~ 9.50817150770749519362447310718916715678773809767440846676787/\ep 
 \nn\\
 &+ 106.846826668085492645539260806397665758174182861898304620986
 \nn\\
 &+ 1086.60508936445510128134626208539389154637371328559182445538 ~ \ep 
 %%%%%%%%%%
 \nn \\
 =&~ \frac{1}{\ep} \bigg( \frac{12 \zeta_2^2}{5}
        +\zeta_2 \zeta_3
        +\zeta_5 \bigg)
 + \bigg( 
 \frac{24}{5} \zeta_2^2
+\frac{111}{10} \zeta_2^3
+16 \zeta_2 \zeta_3
+6 \zeta_3^2
+4 \zeta_5
 \bigg)
 \nn\\
 &+ \ep \bigg( 
 \frac{48}{5} \zeta_2^2
+\frac{2122}{35} \zeta_2^3
+32 \zeta_2 \zeta_3
+\frac{681}{10} \zeta_2^2 \zeta_3
+12 \zeta_3^2
+8 \zeta_5
-\frac{67}{2} \zeta_2 \zeta_5
+\frac{8529}{16} \zeta_7
 \bigg) \,.
% \end{align}
% 
\\
% 
% \begin{align}
 J_3 =&~ 0.00277777777777777777777777777777777777777777777777777777777778/\ep^5
 \nn\\
 &- 0.0361111111111111111111111111111111111111111111111111111111111/\ep^4
 \nn\\
 &+ 0.614342903086572016230704257635362636491570608122828066446491/\ep^3
 \nn\\
 &- 2.34606883344093038863818350703622910763680750137140892258048/\ep^2
 \nn\\
 &+ 20.7239362079225624295936570840634156856361033919705523677456/\ep
 \nn\\
 &- 102.137080500713770454158085984583431631006835833131173671352
 \nn\\
 &+ 707.637488980516140963097357277365579231294612697726106856846 ~ \ep
%%%%%%%%%%%%
 \nn \\
 =& \frac{1}{\ep^5}  \bigg( \frac{1}{360} \bigg)
 + \frac{1}{\ep^4}  \bigg( -\frac{13}{360} \bigg)
 + \frac{1}{\ep^3}  \bigg( \frac{1}{3}+\frac{41}{240} \zeta_2 \bigg)
 + \frac{1}{\ep^2}  \bigg( -\frac{27}{10}
-\frac{133}{240} \zeta_2
+\frac{379}{360} \zeta_3
\bigg)
\nn \\
 &+ \frac{1}{\ep}  \bigg( 
 \frac{102}{5}
+\frac{1}{2} \zeta_2
+\frac{36281 \zeta_2^2}{14400}
-\frac{2191}{360} \zeta_3
 \bigg)
 +   \bigg( 
 -\frac{738}{5}
+\frac{279}{20} \zeta_2
-\frac{95687 \zeta_2^2}{4800}
\nn\\
&+\frac{527}{15} \zeta_3
-\frac{1231}{720} \zeta_2 \zeta_3
+\frac{7253}{200} \zeta_5
 \bigg) 
+ \ep \bigg( 
\frac{5184}{5}
-\frac{927}{5} \zeta_2
+\frac{29703}{200} \zeta_2^2
+\frac{4606121 \zeta_2^3}{120960}
\nn\\
&
-\frac{405}{2} \zeta_3
-\frac{1483}{48} \zeta_2 \zeta_3
+\frac{3529}{144} \zeta_3^2
-\frac{188887}{600} \zeta_5
\bigg)
 \,.
\end{align}

%------------------------------------------------------------------------------------------------------------------

\section{Results} \label{sec:results}
We have computed the three-loop QCD corrections to the HLFF in the color-planar limit.
We have also obtained the necessary contributions from lower orders in $\alpha_s$ i.e. the two-loop QCD corrections to the HLFF up to ${\cal O}(\ep^2)$
and the one-loop QCD corrections to the HLFF up to ${\cal O}(\ep^4)$.
In this section, we present the primary results of this work i.e. the $G_i^{(3)}$s and the $S^{(3)}$.
In order to present them in a concise form, we provide here only the hard finite remainder 
after performing IR subtraction through eq.~\ref{eq:ir}.
$G_i^{(\rm fin)}$ also accepts series expansion in $\alpha_s$ as follows
\begin{equation}
 G_i^{\rm fin} = \sum_{n=0}^\infty \asr^n {\mathcal G}_i^{(n)}  \,.
\end{equation}
We present the renormalized three-loop results (in the color-planar limit) along with the one-loop results up to ${\cal O}(\ep^3)$
and the two-loop results up to ${\cal O}(\ep)$ in the appendices. We also provide an ancillary file \texttt{result.nb}
containing all the results in \textsc{Mathematica} format with the \texttt{arXiv} submission of this manuscript.
We use the mass of the top quark as the renormalization scale i.e. $\mu_R^2 = m_t^2$.
To make the presentation more compact,
we introduce the following abbreviations
for some of the rational functions that appear in the expressions:
\begin{align}
 \bar{x} = \frac{1}{1-x} \,, ~~~
 f_n = \frac{1}{x} - n \,,  ~~~ f_{m,n} = \frac{m}{x} - n \,.  
\end{align}
In the following, we present all the ${\mathcal G}_i^{(3)}$s and ${\mathcal S}^{(3)}$ in the color-planar limit.


\subsection{Checks}
We have performed several checks on our results. 
Firstly, we have cross-checked all our one- and two-loop results with the ones available in literature, finding perfect agreement. Specifically, we have cross-checked all one- and 
two-loop results, both for time-like and space-like virtualities of the boson, 
with the results in the ancillary file \texttt{SemilepFF.txt} included 
in the arXiv submission of ref.~\cite{Bonciani:2008wf}.
As mentioned earlier, we have cross-checked the computation
of the relevant three-loop MIs numerically against \textsc{FIESTA} and \textsc{AMFlow}.
The second and most important check of the UV renormalized three-loop
form factors is that they satisfy the universal IR structure. 
The Ward identity also acts as a non-trivial check 
of our computation. It can easily be checked that the three-loop form factors satisfy eqn.~\ref{eq:WI}.

\section{Asymptotic behaviour of the heavy-light form factors}
\label{sec:asymp}
Scattering amplitudes in massless QCD provide good understanding of the underlying fundamental principles 
such as the factorization or the IR behaviour of the QCD amplitudes.
The massless form factors exponentiate and demonstrate the Sudakov behaviour, which in turn 
has allowed for a rigorous study to predict all the universal contributions of the form factors.
The massive form factors generally do not exponentiate. However, in the asymptotic limit,
when the mass scale is much smaller than the center of mass energy,
they do exponentiate and present a rich structure in terms of IR singularities and high-energy logarithms.
A first study was performed in ref.~\cite{Mitov:2006xs} to comprehend this structure at the two-loop level.
In refs.~\cite{Ahmed:2017gyt,Blumlein:2018tmz}, detailed studies have been performed to find the structures
at three loops and beyond.
The Sudakov behaviour is universal, and hence it is expected that the HLFF will exhibit 
a combined behaviour of the massless and massive form factors. We explicitly study the corresponding
RGE and briefly present our findings in this section.

We begin with the observation in refs.~\cite{Mitov:2006xs,Ahmed:2017gyt,Blumlein:2018tmz} that the massive form factors
factorize into two contributions $\hat{F}_I$ and $\mathcal{C}_I$, as described in the following.
$I$ denotes the corresponding vertex i.e. $I=G_1$ and $S$ for the corresponding form factors. 
The function $\hat{F}_I$, which contains the contributions from the universal logarithms and IR structures,
satisfies the following integro-differential equation 
\begin{equation}
 \mu^2 \frac{\partial}{\partial \mu^2} \ln \hat{F}_I \left( \frac{Q^2}{\mu^2}, \frac{m_t^2}{\mu^2}, \frac{\mu_R^2}{\mu^2}, \alpha_s, \ep \right)
 = \frac{1}{2} \left[ K_I \left( \frac{m_t^2}{\mu^2}, \frac{\mu_R^2}{\mu^2}, \alpha_s, \ep \right) 
                    + G_I \left( \frac{Q^2}{\mu^2}, \frac{\mu_R^2}{\mu^2}, \alpha_s, \ep \right) \right] 
\end{equation}
where $\mu^2$ is the scale introduced for the Sudakov factorization. 
The key factor is that the mass dependence is contained in the function $K_I$, while the process 
dependence through the hard scale $Q^2$ is contained in the $G_I$ function.
The function $\hat{F}_I$ is related to $\tilde{F}_I$, the asymptotic limit of the actual form factors $F_I$, through the matching coefficient ${\mathcal{C}_I}$, 
% which contains all the non-logarithmic contributions to the form factor, 
as follows
\begin{equation}
 \tilde{F}_I \left( \frac{Q^2}{\mu^2}, \frac{m_t^2}{\mu^2}, \frac{\mu_R^2}{\mu^2}, \alpha_s, \ep \right) = 
 \mathcal{C}_I ~ \hat{F}_I \left( \frac{Q^2}{\mu^2}, \frac{m_t^2}{\mu^2}, \frac{\mu_R^2}{\mu^2}, \alpha_s, \ep \right)
 \label{eq:sudakov}
\end{equation}
The $K_I$ and $G_I$ functions also satisfy the RGE which is controlled by the light-like cusp anomalous dimension. 
The solutions of the RGE which involve the respective boundary conditions  
$K_I \left( \alpha_s (m_t^2), 1, \ep \right) \equiv \mathcal{K}_I$ and $G_I \left( \alpha_s (Q^2), 1, \ep \right) \equiv \mathcal{G}_I$,
and eq.~\ref{eq:sudakov} provide the $\hat{F}_I$ in terms of these finite functions ($\mathcal{K}_I$, $\mathcal{G}_I$)
and the light-like cusp anomalous dimension ($A_q$ in the present scenario).
The solutions of $\hat{F}_I$ in terms of $\mathcal{K}_I$, $\mathcal{G}_I$ and $A_I$ up to four-loop are presented in ref.~\cite{Blumlein:2018tmz} for the HQFF.
We note that these solutions do not contain the contributions from massive internal quark loops.
The corresponding solutions for the HLFF are same except the fact that the $\mathcal{K}_I$ have to be interpreted differently as 
described in the following, and the Sudakov logarithm is defined as followed
\begin{equation}
 L = \ln \left(- \frac{q^2}{m_t^2} \right)
\end{equation}
$\mathcal{K}_I$ encapsulate the universality of the IR structures, and hence we anticipate that 
$\mathcal{K}_I$ in the present case will show a combined behaviour of the respective functions $\mathcal{K}_{I,0}$ and $\mathcal{K}_{I,m}$
in the massless and massive form factors. 
Indeed, we find
\begin{equation}
\label{eq:KI}
 \mathcal{K}_I = \frac{1}{2} (\mathcal{K}_{I,0} + \mathcal{K}_{I,m}) \,.
\end{equation}
Eqs.~(10)-(13) in ref.~\cite{Blumlein:2018tmz} along with eq.~\eqref{eq:KI} provide the $\hat{F}_I$ for the HLFF.
All the appearing anomalous dimensions admit series expansion in $\alpha_s$.
The light-like cusp anomalous dimension is now known up to four-loop order \cite{Moch:2004pa,Vogt:2004mw,Henn:2019swt,vonManteuffel:2020vjv}.
$\mathcal{K}_{I,0}$ has been computed in terms of the light-like cusp anomalous dimension and presented up to four loops in \cite{Ravindran:2005vv}. 
In \cite{Blumlein:2018tmz}, $\mathcal{K}_{I,m}$ has been conjectured to be a combined contributions from the collinear ($B_I$), soft ($f_I$),
quark mass ($\gamma_m$) and heavy quark ($\gamma_Q$) anomalous dimensions. 
$\gamma_q = B_q + \frac{f_q}{2}$ is available up to four-loop level \cite{Moch:2004pa,Vogt:2004mw,Henn:2019swt,vonManteuffel:2020vjv}. 
$\gamma_m$ is also available up to five-loop level \cite{Broadhurst:1991fy,Vermaseren:1997fq,Melnikov:2000zc,Marquard:2007uj,Luthe:2016xec,Baikov:2017ujl}
whereas
$\gamma_Q$ is known up to three-loop level \cite{Becher:2009kw,Blumlein:2018tmz}.

In summary, all the constituents in the generic structure of $\hat{F}_I$ are known except for the other finite function $\mathcal{G}_I$.
At this point, we consider our exact computation of the HLFF up to two-loop in the asymptotic limit, and compare
order by order in $\ep$ and $L$.
We find perfect agreement for the poles in $\ep$ except for the single pole.
In the next, by demanding the equality again order by order in $\ep$ and $L$, we find the finite function $\mathcal{G}_I$ 
and the matching coefficient $\mathcal{C}_I$.
The finite function $\mathcal{G}_I$ captures the $Q^2$ dependence of the form factors, not the mass dependence,
and hence we find them the same as in the case of massless form factors, as expected.
We also find that the matching coefficients of HLFF and HQFF ($\mathcal{C}_{I,m}$) also are related. In case of the vector form factor,
\begin{equation}
 \mathcal{C}_V = \mathcal{C}_{V,m}^{\scriptstyle \frac{1}{2}} \,.
\end{equation}
The square-root is also expected as each massive leg contributes equally to the matching coefficient, and 
in case of the HLFF, there is single massive leg present.
However in case of the scalar form factors, the overall renormalization of the Yukawa coupling are the same
in both the HLFF and the HQFF. Hence, $\mathcal{C}_S$ contains an additional $Z_m^{\frac{1}{2}}$.

Next, we extend our study to the higher orders in $\alpha_s$. Now, we use all the available quantities, and 
predict the three- and four-loop HLFF in the asymptotic limit. For the three-loop form factors, we
predict all the poles and also the complete logarithmic contributions to the finite part
($\tilde{G}_1^{(3,0)}$ and $\tilde{S}^{(3,0)}$ : the $\ep^0$ coefficient of the three-loop form factors $G_1^{(3)}$ and $S^{(3)}$
in the asymptotic limit).
The three-loop matching coefficient is known partially i.e. the complete light-quark and the color-planar contributions.
Hence, the non-logarithmic contributions to $\tilde{G}_1^{(3,0)}$ and $\tilde{S}^{(3,0)}$ have been obtained 
for these color configurations.
We take the color-planar contributions of the predicted results and compare those with our computed ones
after performing appropriate Taylor series expansion in the large-$x$ limit, finding complete agreement.
This provides strong checks on both the computations. On one hand, it ensures that the three-loop computation
is correct and on the other hand, it showcases the richness of the universal structure of the QCD amplitudes.

% % 
In the following, we present $\tilde{G}_1^{(3,0)}$ and $\tilde{S}^{(3,0)}$.
We define the following two constants 
\begin{align}
 c_1 &= 12 \zeta_2 \log ^2(2) + \log^4(2) + 24 \text{Li}_4 \Big( \frac{1}{2}\Big) \,,
 \\
 c_2 &= 26 \zeta_2^2 \log (2) - 20 \zeta_2 \log ^3(2) - \log ^5(2) +120 \text{Li}_5 \Big( \frac{1}{2} \Big) \,.
\end{align}
We present all the logarithmic contributions and the partial non-logarithmic contributions originating from the complete light-quark
and color-planar corrections.
% 
% 
% % 
% 
% 
\begin{align}
 \tilde{G}_1^{(3,0)} &= 
 C_F^3 \bigg[
-\frac{169}{144} L^6
+\frac{2473}{240} L^5
+L^4 \bigg(
        -\frac{6379}{96}
        +\frac{97 \zeta_2}{16}
\bigg)
+L^3 \bigg(
        \frac{12689}{48}
        +\frac{75 \zeta_2}{8}
        -\frac{715 \zeta_3}{6}
\bigg)
\nn\\
& ~~~
+L^2 \bigg(
        -\frac{5643}{8}
        -\frac{2331}{16} \zeta_2
        +\frac{6211}{80} \zeta_2^2
        +\frac{1907}{4} \zeta_3
        +24 \zeta_2 \log (2)
\bigg)
+L \bigg(
        \frac{8361}{8}
\nn\\
& ~~~        
        +\frac{10057}{16} \zeta_2
        -\frac{24107}{80} \zeta_2^2
        -\frac{5161}{4} \zeta_3
        +\frac{209}{2} \zeta_2 \zeta_3
        -\frac{1623}{5} \zeta_5
        -96 \zeta_2 \log (2)
        +8 c_1
\bigg)
 \bigg]
% % % % % % % % % % % % % % % % % % % % % % %  
\nn\\&
+ C_A C_F^2 \bigg[
-\frac{451}{144} L^5
+L^4 \bigg(
        \frac{39395}{864}
        -\frac{145 \zeta_2}{24}
\bigg)
+L^3 \bigg(
        -\frac{61127}{216}
        -\frac{709 \zeta_2}{72}
        +\frac{583 \zeta_3}{6}
\bigg)
\nn\\
& ~~~
+L^2 \bigg(
        \frac{709445}{648}
        +\frac{6589}{48} \zeta_2
        -\frac{1033}{20} \zeta_2^2
        -524 \zeta_3
        -12 \zeta_2 \log (2)
\bigg)
+L \bigg(
        -\frac{482273}{216}
\nn\\
& ~~~
        -\frac{5117}{8} \zeta_2
        +\frac{60547}{240} \zeta_2^2
        +\frac{22417}{12} \zeta_3
        -\frac{261}{2} \zeta_2 \zeta_3
        +116 \zeta_5
        +48 \zeta_2 \log (2)
        -4 c_1
\bigg)
\bigg]
% % % % % % % % % % % % % % % % % % % % % % % % % % % 
\nn\\&
+ C_A^2 C_F \bigg[
-\frac{121}{54} L^4
+L^3 \bigg(
        \frac{2869}{81}
        -\frac{44}{9} \zeta_2
\bigg)
+L^2 \bigg(
        -\frac{18682}{81}
        +\frac{26}{9} \zeta_2
        -\frac{44}{5} \zeta_2^2
        +88 \zeta_3
\bigg)
\nn\\
& ~~~
+L \bigg(
        \frac{1045955}{1458}
        +\frac{17366}{81} \zeta_2
        -\frac{94}{3} \zeta_2^2
        -\frac{17464}{27} \zeta_3
        +\frac{88}{3} \zeta_2 \zeta_3
        +136 \zeta_5
\bigg)
\bigg]
% % % % % % % % % % % % % % % % % % % % % 
\nn\\&
+ C_F n_l^2 T_F^2 \bigg[
 \frac{304}{81} L^3
-\frac{8}{27} L^4
+L^2 \bigg(
        -\frac{1624}{81}
        -\frac{32 \zeta_2}{9}
\bigg)
+L \bigg(
        \frac{39352}{729}
        +\frac{608 \zeta_2}{27}
        +\frac{64 \zeta_3}{27}
\bigg)
\nn\\& ~~~
-\frac{322979}{6561}
-\frac{712}{27} \zeta_2
+\frac{1016}{135} \zeta_2^2
+\frac{2624}{243} \zeta_3
\bigg]
% % % % % % % % % % % % % % % % % % % % % 
% % % % % % % % % % % % % % % % % % % % % 
\nn\\&
+ C_F^2 n_l T_F \bigg[
 \frac{41}{36} L^5
-\frac{3265}{216} L^4
+L^3 \bigg(
        \frac{4987}{54}
        +\frac{149 \zeta_2}{18}
\bigg)
+L^2 \bigg(
        -\frac{53683}{162}
        -\frac{599 \zeta_2}{12}
        +29 \zeta_3
\bigg)
\nn\\
& ~~~
+L \bigg(
        \frac{32273}{54}
        +\frac{345}{2} \zeta_2
        -\frac{389}{60} \zeta_2^2
        -\frac{2153}{9} \zeta_3
\bigg)
-\frac{1132067}{2916}
-\frac{16211}{54} \zeta_2
+\frac{349843 \zeta_2^2}{2160}
\nn\\& ~~~
+\frac{72127}{162} \zeta_3
+\frac{25}{18} \zeta_2 \zeta_3
+\frac{229}{9} \zeta_5
+\frac{224}{3} \zeta_2 \log (2)
-\frac{32 c_1}{9}
\bigg]
% 
% % % % % % % % % % % % % % % % % % % % % 
\nn\\&
+ C_F C_A n_l T_F \bigg[
 \frac{44}{27} L^4
+L^3 \bigg(
        -\frac{1948}{81}
        +\frac{16 \zeta_2}{9}
\bigg)
+L^2 \bigg(
        \frac{11752}{81}
        +\frac{32 \zeta_2}{3}
        -16 \zeta_3
\bigg)
\nn\\
& ~~~
+L \bigg(
        -\frac{309838}{729}
        -\frac{11728}{81} \zeta_2
        +\frac{88}{15} \zeta_2^2
        +\frac{1448}{9} \zeta_3
\bigg)
+\frac{2866346}{6561}
+\frac{129091}{729} \zeta_2
\nn\\
& ~~~
-\frac{10922}{135} \zeta_2^2
-\frac{28424}{81} \zeta_3
+\frac{8}{3} \zeta_2 \zeta_3
+98 \zeta_5
-\frac{112}{3} \zeta_2 \log (2)
+\frac{16 c_1}{9}
\bigg]
% 
% % % % % % % % % % % % % % % % % % 
\nn\\
&
+ N_C^3 \bigg[
% 
% \big(
        -\frac{1599187}{104976}
        -\frac{550147}{46656} \zeta_2
        -\frac{3286613}{69120} \zeta_2^2
        +\frac{5780057}{241920} \zeta_2^3
        +\frac{2842675}{31104} \zeta_3
\nn\\
& ~~~
        +\frac{8099}{576} \zeta_2 \zeta_3
        -\frac{7}{288} \zeta_3^2
        -\frac{6281}{32} \zeta_5
% \big)
% 
\bigg]
+ X_{G_1}^{(3,0)} \,.
\\
% % % % % % % % % % % % % % % % % % % % % % % % % %      S30
 \tilde{S}^{(3,0)} &= 
 C_F^3 \bigg[
-\frac{169}{144} L^6
+\frac{31}{24} L^5
+L^4 \bigg(
        -\frac{1309}{96}
        +\frac{97}{16} \zeta_2
\bigg)
+L^3 \bigg(
        \frac{1177}{24}
        +\frac{165}{4} \zeta_2
        -\frac{715}{6} \zeta_3
\bigg)
\nn\\& ~~~
+L^2 \bigg(
        -\frac{2817}{16}
        -\frac{1549}{16} \zeta_2
        +\frac{6211}{80} \zeta_2^2
        +\frac{655}{2} \zeta_3
        -24 \zeta_2 \log (2)
\bigg)
+L \bigg(
        \frac{3577}{16}
        +\frac{321}{8} \zeta_2
\nn\\& ~~~
        -\frac{1531}{40} \zeta_2^2
        -\frac{3655}{4} \zeta_3
        +\frac{209}{2} \zeta_2 \zeta_3
        -\frac{1623}{5} \zeta_5
        +552 \zeta_2 \log (2)
        -8 c_1
\bigg)
\bigg]
\nn\\&
+ C_A C_F^2 \bigg[
-\frac{451}{144} L^5
+L^4 \bigg(
        \frac{2635}{108}
        -\frac{145}{24} \zeta_2
\bigg)
+L^3 \bigg(
        -\frac{12737}{216}
        -\frac{1375}{72} \zeta_2
        +\frac{583}{6} \zeta_3
\bigg)
\nn\\& ~~~
+L^2 \bigg(
        \frac{235093}{1296}
        +\frac{613}{12} \zeta_2
        -\frac{1033}{20} \zeta_2^2
        -\frac{1271}{4} \zeta_3
        +12 \zeta_2 \log (2)
\bigg)
+L \bigg(
        -\frac{155401}{432}
\nn\\& ~~~
        -\frac{6121}{24} \zeta_2
        +\frac{23503}{240} \zeta_2^2
        +\frac{2507}{2} \zeta_3
        -\frac{261}{2} \zeta_2 \zeta_3
        +116 \zeta_5
        -276 \zeta_2 \log (2)
        +4 c_1
\bigg)
\bigg]
\nn\\&
+ C_A^2 C_F  \bigg[
-\frac{121}{54} L^4
+L^3 \bigg(
        \frac{1780}{81}
        -\frac{44}{9} \zeta_2
\bigg)
+L^2 \bigg(
        -\frac{11939}{162}
        +\frac{26}{9} \zeta_2
        -\frac{44}{5} \zeta_2^2
        +88 \zeta_3
\bigg)
\nn\\& ~~~
+L \bigg(
        \frac{10289}{1458}
        +\frac{9644}{81} \zeta_2
        -\frac{94}{3} \zeta_2^2
        -\frac{13900}{27} \zeta_3
        +\frac{88}{3} \zeta_2 \zeta_3
        +136 \zeta_5
\bigg)
\bigg]
\nn\\&
+ C_F n_l^2 T_F^2  \bigg[
-\frac{8}{27} L^4
+\frac{160}{81} L^3
+L^2 \bigg(
        -\frac{400}{81}
        -\frac{32}{9} \zeta_2
\bigg)
+L \bigg(
        \frac{3712}{729}
        +\frac{320}{27} \zeta_2
        +\frac{64}{27} \zeta_3
\bigg)
\nn\\& ~~~
-\frac{59891}{6561}
-\frac{1144}{27} \zeta_2
+\frac{1016}{135} \zeta_2^2
-\frac{2560}{243} \zeta_3
\bigg]
\nn\\&
+ C_F^2 n_l T_F  \bigg[
 \frac{41}{36} L^5
-\frac{200}{27} L^4
+L^3 \bigg(
        \frac{1213}{54}
        +\frac{149}{18} \zeta_2
\bigg)
+L^2 \bigg(
        -\frac{19625}{324}
        -\frac{83}{3} \zeta_2
        +29 \zeta_3
\bigg)
\nn\\&~~~
+L \bigg(
        \frac{13309}{108}
        +\frac{809}{6} \zeta_2
        -\frac{389}{60} \zeta_2^2
        -\frac{1994}{9} \zeta_3
\bigg)
-\frac{195257}{2916}
-\frac{343}{27} \zeta_2
+\frac{6547}{2160} \zeta_2^2
\nn\\&~~~
+\frac{64171}{162} \zeta_3
+\frac{25}{18} \zeta_2 \zeta_3
+\frac{229}{9} \zeta_5
-\frac{1184}{3} \zeta_2 \log (2)
+\frac{32 c_1}{9}
\bigg]
\nn\\&
+ C_A C_F n_l T_F  \bigg[
 \frac{44}{27} L^4
+L^3 \bigg(
        -\frac{1156}{81}
        +\frac{16}{9} \zeta_2
\bigg)
+L^2 \bigg(
        \frac{3454}{81}
        +\frac{32}{3} \zeta_2
        -16 \zeta_3
\bigg)
\nn\\&~~~
+L \bigg(
        -\frac{14998}{729}
        -\frac{6544}{81} \zeta_2
        +\frac{88}{15} \zeta_2^2
        +\frac{1448}{9} \zeta_3
\bigg)
+\frac{854630}{6561}
+\frac{77899}{729} \zeta_2
\nn\\&~~~
-\frac{9158}{135} \zeta_2^2
-\frac{16652}{81} \zeta_3
+\frac{8}{3} \zeta_2 \zeta_3
+98 \zeta_5
+\frac{592}{3} \zeta_2 \log (2)
-\frac{16 c_1}{9}
\bigg]
\nn\\&
+ N_C^3  \bigg[
\frac{4233773}{419904}
-\frac{1901551}{46656} \zeta_2
-\frac{125401}{13824} \zeta_2^2
+\frac{5780057}{241920} \zeta_2^3
+\frac{397231}{31104} \zeta_3
+\frac{22571}{576} \zeta_2 \zeta_3
\nn\\&~~~
-\frac{7}{288} \zeta_3^2
-\frac{4829}{32} \zeta_5
\bigg]
+ X_{S}^{(3,0)} \,.
\end{align}
% 
% % 

% 
% 
The $X_{G_1}^{(3,0)}$ and $X_{S}^{(3,0)}$ denote the color-non-planar contributions to the non-logarithmic part of $\tilde{G}_1^{(3,0)}$ and
$\tilde{S}^{(3,0)}$, respectively.
In the appendix \ref{sec:res4L}, we also present the four-loop form factors ($\tilde{G}_1^{(4)}$ and $\tilde{S}^{(4)}$) in the asymptotic limit.
The available ingredients allow us to compute up to the $\frac{1}{\ep^3}$ pole with complete color dependence.
For the complete light-quark contributions, we obtain complete result for the $\frac{1}{\ep^2}$ pole,
complete logarithmic result for the $\frac{1}{\ep}$ pole and up to $\mathcal{O}(L^2)$ for the finite part.
For the rest of the contributions, we obtain one less order in $L$ compared to the light-quark contributions for each $\ep$ order.

\section{Conclusions} \label{sec:conclusions}
In this paper, we have presented the three-loop QCD corrections to the heavy-light 
form factors in the color-planar limit. We have used the generic techniques 
of IBP reduction and the method of differential equations to compute the 
bare form factors. 
Following the appropriate ultraviolet renormalization procedure through a mixed scheme, 
we obtain the renormalized form factors which satisfy the universal infrared behaviour.
The form factors are expressed in terms of harmonic polylogarithms 
and generalized harmonic polylogarithms. 
The results presented in this paper are the first steps to obtain 
a precise third-order QCD corrections to the heavy-to-light quark transition.
The results will be very important in the precise determination of the 
properties of the top quark decay at third order in the strong coupling 
constant. The results can also be combined with the jet and soft function
in the SCET framework to obtain a precise phenomenological determination 
of the CKM matrix elements $|V_{cb}|$ and $|V_{ub}|$. Hence, they are
also very important for the study of flavour and CP violation
within and beyond the SM of particle physics.

We have also studied the asymptotic behaviour of the HLFF by solving 
the integro-differential Sudakov equation and the corresponding RGE.
The systematic study has allowed us to find a connection among all types of form factors
and also enabled us to obtain complete logarithmic corrections at three–loop level.
We also compute partial four–loop results, which can be fully determined by performing an explicit four–loop calculation
of one of these massive form factors.

\section*{Acknowledgements}
We would like to thank B. Ananthanarayan, J. Bl\"umlein, R. Bonciani, P. Marquard
and A. Vicini for fruitful discussions and their comments on the manuscript.
We specially thank S. Banik for being part of the initial collaboration and
for his support with the package \textsc{MBConicHulls}.
We also thank A. Sankar for providing the necessary ingredients from massless form factors.

\appendix

\section{One-loop form factors}
\label{sec:res1L}
In this appendix, we present the one-loop form factors up to ${\cal O}(\ep^3)$.
\begin{align}
 G_1^{(1)} &= C_F \Big[
 \frac{1}{\ep^2} \Big\{  
 -1
 \Big\}
+  \frac{1}{\ep} \Big\{ 
-\frac{5}{2}-2 H_1(x)
 \Big\}
+ \Big\{  
-6
+\fxthr H_1(x)
-2 H_{0,1}(x)
-4 H_{1,1}(x)
\nn\\&
-\frac{1}{2} \zeta_2
 \Big\}
+ \ep \Big\{  
-12
+\fxthr \big(
        H_{0,1}(x)
        +2 H_{1,1}(x)
\big)
+4 \fxt H_1(x)
-2 H_{0,0,1}(x)
-4 H_{0,1,1}(x)
\nn\\&
-4 H_{1,0,1}(x)
-8 H_{1,1,1}(x)
-\frac{5}{4} \zeta_2
-H_1(x) \zeta_2
+\frac{1}{3} \zeta_3
 \Big\}
+ \ep^2 \Big\{  
-24
+\fxt \big(
        8 H_1(x)
\nn\\&        
        +4 H_{0,1}(x)
        +8 H_{1,1}(x)
\big)
+\fxthr \big(
        2 H_{0,1,1}(x)
        +2 H_{1,0,1}(x)
        +4 H_{1,1,1}(x)
        +\frac{1}{2} H_1(x) \zeta_2
\nn\\&        
        +H_{0,0,1}(x)
\big)
-2 H_{0,0,0,1}(x)
-4 H_{0,0,1,1}(x)
-4 H_{0,1,0,1}(x)
-8 H_{0,1,1,1}(x)
-4 H_{1,0,0,1}(x)
\nn\\&
-8 H_{1,0,1,1}(x)
-8 H_{1,1,0,1}(x)
-16 H_{1,1,1,1}(x)
-3 \zeta_2
-H_{0,1}(x) \zeta_2
-2 H_{1,1}(x) \zeta_2
-\frac{9}{40} \zeta_2^2
\nn\\&
+\frac{5}{6} \zeta_3
+\frac{2}{3} H_1(x) \zeta_3
 \Big\}
+ \ep^3 \Big\{  
-48
+\fxt \big(
        16 H_1(x)
        +8 H_{0,1}(x)
        +16 H_{1,1}(x)
        +4 H_{0,0,1}(x)
\nn\\&        
        +8 H_{0,1,1}(x)
        +8 H_{1,0,1}(x)
        +16 H_{1,1,1}(x)
        +2 H_1(x) \zeta_2
\big)
+\fxthr \big(
        2 H_{0,0,1,1}(x)
        +2 H_{0,1,0,1}(x)
\nn\\&        
        +4 H_{0,1,1,1}(x)
        +2 H_{1,0,0,1}(x)
        +4 H_{1,0,1,1}(x)
        +4 H_{1,1,0,1}(x)
        +8 H_{1,1,1,1}(x)
\nn\\&        
        +\big(
                \frac{1}{2} H_{0,1}(x)
                +H_{1,1}(x)
        \big) \zeta_2
        -\frac{1}{3} H_1(x) \zeta_3
        +H_{0,0,0,1}(x)
\big)
-2 H_{0,0,0,0,1}(x)
-4 H_{0,0,0,1,1}(x)
\nn\\&
-4 H_{0,0,1,0,1}(x)
-8 H_{0,0,1,1,1}(x)
-4 H_{0,1,0,0,1}(x)
-8 H_{0,1,0,1,1}(x)
-8 H_{0,1,1,0,1}(x)
\nn\\&
-16 H_{0,1,1,1,1}(x)
-4 H_{1,0,0,0,1}(x)
-8 H_{1,0,0,1,1}(x)
-8 H_{1,0,1,0,1}(x)
-16 H_{1,0,1,1,1}(x)
\nn\\&
-8 H_{1,1,0,0,1}(x)
-16 H_{1,1,0,1,1}(x)
-16 H_{1,1,1,0,1}(x)
-32 H_{1,1,1,1,1}(x)
+\big(
        -6
        -H_{0,0,1}(x)
\nn\\&       
        -2 H_{0,1,1}(x)
        -2 H_{1,0,1}(x)
        -4 H_{1,1,1}(x)
        +\frac{1}{6} \zeta_3
\big) \zeta_2
+\big(
        2
        +\frac{2}{3} H_{0,1}(x)
        +\frac{4}{3} H_{1,1}(x)
\big) \zeta_3
\nn\\&
+\frac{1}{5} \zeta_5
-\frac{9}{80} \zeta_2^2 \big(
        5+4 H_1(x)\big)
 \Big\}
%  
% + \ep^4 \Big\{  
%  \Big\}
%  
 \Bigg] \,.
\\
% % % % % % % % % % % % % % % % % % % % % % % % % % % % 
 G_2^{(1)} &= C_F \bigg[
%  \frac{1}{\ep^2} \Big\{  
%  \Big\}
% %  
% +  \frac{1}{\ep} \Big\{  
%  \Big\}
%  
% + 
\Big\{
-\frac{2 H_1(x)}{x}
 \Big\}
- \ep \Big\{  
\frac{2}{x} \big(
        4 H_1(x)
        +2 H_{1,1}(x)
        +H_{0,1}(x)
\big)
 \Big\}
+ \ep^2 \Big\{  
-\frac{1}{x} \big(
        2 \big(
                4 H_{0,1}(x)
\nn\\&                
                +8 H_{1,1}(x)
                +2 H_{0,1,1}(x)
                +2 H_{1,0,1}(x)
                +4 H_{1,1,1}(x)
                +H_{0,0,1}(x)
        \big)
        +\big(
                16
                +\zeta_2
        \big) H_1(x)
\big)
 \Big\}
\nn\\&
+ \ep^3 \Big\{  
-\frac{1}{3 x} \big(
        6 \big(
                \big(
                        16
                        +\zeta_2
                \big) H_{1,1}(x)
                +4 H_{0,0,1}(x)
                +8 H_{0,1,1}(x)
                +8 H_{1,0,1}(x)
                +16 H_{1,1,1}(x)
\nn\\&                
                +2 H_{0,0,1,1}(x)
                +2 H_{0,1,0,1}(x)
                +4 H_{0,1,1,1}(x)
                +2 H_{1,0,0,1}(x)
                +4 H_{1,0,1,1}(x)
                +4 H_{1,1,0,1}(x)
\nn\\&                
                +8 H_{1,1,1,1}(x)
                +H_{0,0,0,1}(x)
        \big)
        +2 \big(
                48
                +6 \zeta_2
                -\zeta_3
        \big) H_1(x)
        +3 \big(
                16
                +\zeta_2
        \big) H_{0,1}(x)
\big)
 \Big\}
%  
% + \ep^4 \Big\{  
%  \Big\}
%  
 \bigg] \,.
\\
 G_3^{(1)} &= C_F \bigg[
%  \frac{1}{\ep^2} \Big\{  
%  \Big\}
%  
% +  \frac{1}{\ep} \Big\{  
%  \Big\}
%  
% + 
\Big\{ 
\frac{4-2 \fxtnt H_1(x)}{x}
 \Big\}
+  \frac{\ep}{x} \Big\{
8
+\fxtnt \big(
        -2 H_{0,1}(x)
        -4 H_{1,1}(x)
\big)
-4 \fxthr H_1(x)
 \Big\}
\nn\\&
+  \frac{\ep^2}{x} \Big\{  
16
+\fxthr \big(
        -8 H_1(x)
        -4 H_{0,1}(x)
        -8 H_{1,1}(x)
\big)
+\fxtnt \big(
        -2 H_{0,0,1}(x)
        -4 H_{0,1,1}(x)
\nn\\&        
        -4 H_{1,0,1}(x)
        -8 H_{1,1,1}(x)
        -H_1(x) \zeta_2
\big)
+2 \zeta_2
 \Big\}
+ \frac{\ep^3}{x} \Big\{  
32
+\fxthr \big(
        -16 H_1(x)
        -8 H_{0,1}(x)
\nn\\&        
        -16 H_{1,1}(x)
        -4 H_{0,0,1}(x)
        -8 H_{0,1,1}(x)
        -8 H_{1,0,1}(x)
        -16 H_{1,1,1}(x)
        -2 H_1(x) \zeta_2
\big)
\nn\\&
+\fxtnt \big(
        -2 H_{0,0,0,1}(x)
        -4 H_{0,0,1,1}(x)
        -4 H_{0,1,0,1}(x)
        -8 H_{0,1,1,1}(x)
        -4 H_{1,0,0,1}(x)
\nn\\&        
        -8 H_{1,0,1,1}(x)
        -8 H_{1,1,0,1}(x)
        -16 H_{1,1,1,1}(x)
        +\big(
                -H_{0,1}(x)
                -2 H_{1,1}(x)
        \big) \zeta_2
\nn\\&        
        +\frac{2}{3} H_1(x) \zeta_3
\big)
+4 \zeta_2
-\frac{4}{3} \zeta_3
 \Big\}
%  
% + \ep^4 \Big\{  
%  \Big\}
%  
 \bigg]\,.
\\
% % % % % % % % % % % % % % % % % % % % % % 
 S^{(1)} &= C_F \bigg[
 \frac{1}{\ep^2} \Big\{  
 -1
 \Big\}
+  \frac{1}{\ep} \Big\{  
-\frac{5}{2}-2 H_1(x)
 \Big\}
+ 
\Big\{ 
-4
-2 \ixo H_1(x)
-2 H_{0,1}(x)
-4 H_{1,1}(x)
\nn\\&
-\frac{1}{2} \zeta_2
 \Big\}
+  \ep \Big\{
-8
-2 \fxmo H_1(x)
-2 \ixo H_{0,1}(x)
-4 \ixo H_{1,1}(x)
-2 H_{0,0,1}(x)
\nn\\&
-4 H_{0,1,1}(x)
-4 H_{1,0,1}(x)
-8 H_{1,1,1}(x)
+\frac{1}{3} \zeta_3
-\frac{1}{4} \zeta_2 \big(
         5+4 H_1(x)\big)
 \Big\}
\nn\\&
+  \ep^2 \Big\{  
-16
+\fxmo \big(
        -4 H_1(x)        
        -2 H_{0,1}(x)
        -4 H_{1,1}(x)
\big)
-2 \ixo H_{0,0,1}(x)
\nn\\&
-4 \ixo H_{0,1,1}(x)
-4 \ixo H_{1,0,1}(x)
-8 \ixo H_{1,1,1}(x)
-2 H_{0,0,0,1}(x)
-4 H_{0,0,1,1}(x)
\nn\\&
-4 H_{0,1,0,1}(x)
-8 H_{0,1,1,1}(x)
-4 H_{1,0,0,1}(x)
-8 H_{1,0,1,1}(x)
\nn\\&
-8 H_{1,1,0,1}(x)
-16 H_{1,1,1,1}(x)
+\big(
        -2
        -\ixo H_1(x)
        -H_{0,1}(x)
        -2 H_{1,1}(x)
\big) \zeta_2
-\frac{9}{40} \zeta_2^2
\nn\\&
+\frac{1}{6} \zeta_3 \big(
        5+4 H_1(x)\big)
 \Big\}
+ \ep^3 \Big\{  
-32
+\fxmo \big(
        -8 H_1(x)
        -4 H_{0,1}(x)
        -8 H_{1,1}(x)
\nn\\&        
        -2 H_{0,0,1}(x)
        -4 H_{0,1,1}(x)
        -4 H_{1,0,1}(x)
        -8 H_{1,1,1}(x)
\big)
-2 \ixo H_{0,0,0,1}(x)
\nn\\&
-4 \ixo H_{0,0,1,1}(x)
-4 \ixo H_{0,1,0,1}(x)
-8 \ixo H_{0,1,1,1}(x)
-4 \ixo H_{1,0,0,1}(x)
\nn\\&
-8 \ixo H_{1,0,1,1}(x)
-8 \ixo H_{1,1,0,1}(x)
-16 \ixo H_{1,1,1,1}(x)
-2 H_{0,0,0,0,1}(x)
\nn\\&
-4 H_{0,0,0,1,1}(x)
-4 H_{0,0,1,0,1}(x)
-8 H_{0,0,1,1,1}(x)
-4 H_{0,1,0,0,1}(x)
-8 H_{0,1,0,1,1}(x)
\nn\\&
-8 H_{0,1,1,0,1}(x)
-16 H_{0,1,1,1,1}(x)
-4 H_{1,0,0,0,1}(x)
-8 H_{1,0,0,1,1}(x)
-8 H_{1,0,1,0,1}(x)
\nn\\&
-16 H_{1,0,1,1,1}(x)
-8 H_{1,1,0,0,1}(x)
-16 H_{1,1,0,1,1}(x)
-16 H_{1,1,1,0,1}(x)
-32 H_{1,1,1,1,1}(x)
\nn\\&
+\big(
        -4
        +(-1-\ixo) H_1(x)
        -\ixo H_{0,1}(x)
        -2 \ixo H_{1,1}(x)
        -H_{0,0,1}(x)
        -2 H_{0,1,1}(x)
\nn\\&        
        -2 H_{1,0,1}(x)
        -4 H_{1,1,1}(x)
        +\frac{1}{6} \zeta_3
\big) \zeta_2
+\big(
        \frac{4}{3}
        +\frac{2}{3} \ixo H_1(x)
        +\frac{2}{3} H_{0,1}(x)
        +\frac{4}{3} H_{1,1}(x)
\big) \zeta_3
\nn\\&
+\frac{1}{5} \zeta_5
-\frac{9}{80} \zeta_2^2 \big(
        5+4 H_1(x)\big)
 \Big\}
%  
% + \ep^4 \Big\{  
%  \Big\}
%  
 \bigg]\,.
\end{align}

\section{Two-loop form factors}
\label{sec:res2L}
The two-loop form factors are presented up to ${\cal O}(\ep)$ in the following.


\section{Three-loop form factors}
\label{sec:res3L}
The form factor $G_1^{(3)}$ in the color-planar limit, is given by
%
The form factor $G_2^{(3)}$ in the color-planar limit, is given by
\begin{align}
G_2^{(3)} &= 
N_C^3 \bigg[
%  \frac{1}{\ep^6} \Big\{  
%  \Big\}
% %  
% +  \frac{1}{\ep^5} \Big\{  
%  \Big\}
%  
% + 
\frac{1}{\ep^4} \Big\{  
-\frac{1}{8} \ixo H_1(x)
 \Big\}
+  \frac{1}{\ep^3} \Big\{  
-\frac{5}{2} \ixo H_1(x)
-\frac{1}{8} \ixo H_{0,1}(x)
-\frac{5}{4} \ixo H_{1,1}(x)
 \Big\}
\nn\\&
+  \frac{1}{\ep^2} \Big\{  
\iomx^4 \big(
        6+37 x+20 x^2
\big)
\big(2 H_{0,0,1,1}(x)
        -H_{0,1,0,1}(x)
        -3 H_{0,1}(x) \zeta_2
        +\frac{27}{10} \zeta_2^2
\nn\\&        
        +H_{0,0,0,1}(x)
\big)
+\iomx \big(
        \frac{3}{2}
        +\frac{1}{288} (2509-1069 \ixo) H_1(x)
\big)
+\iomx^3 \big(
        \frac{1}{12} \big(
                189
                -5 \ixo
\nn\\&                
                -765 x
                -175 x^2
        \big) H_{0,1}(x)
        +\frac{1}{8} \big(
                151
                +3 \ixo
                +333 x
                +17 x^2
        \big) H_{0,0,1}(x)
        +\frac{1}{4} \big(
                187
                -9 \ixo
\nn\\&                
                +297 x
                +29 x^2
        \big) H_{0,1,1}(x)
        +\frac{1}{4} \big(
                -71
                -3 \ixo
                -171 x
                -7 x^2
        \big) H_{1,0,1}(x)
        +\big(
                -28
                -138 x
\nn\\&                
                -23 x^2
                -\frac{3}{16} \big(
                        305
                        +5 \ixo
                        +663 x
                        +35 x^2
                \big) H_1(x)
        \big) \zeta_2
\big)
+\iomx^2 (73
-12 \ixo
+2 x
) H_{1,1}(x)
\nn\\&
-\frac{19}{2} \ixo H_{1,1,1}(x)
 \Big\}
+  \frac{1}{\ep} \Big\{  
\iomx \big(
        \frac{49}{4}
        +\frac{1}{864} (-2527+37519 \ixo) H_1(x)
\big)
\nn\\&
+\iomx^2 \big(
        \frac{1}{48} \big(
                10894
                +841 \ixo
                +24 \ixtw
                +1993 x
        \big) H_{1,1}(x)
        +(656
        -34 \ixo
        +134 x
        ) H_{1,1,1}(x)
\big)
\nn\\&
+\iomx^4 \big(
        \big(
                6+37 x+20 x^2
        \big)
\big(4 H_{0,0,0,0,1}(x)
                +12 H_{0,0,0,1,1}(x)
                +7 H_{0,0,1,0,1}(x)
\nn\\&                
                +24 H_{0,0,1,1,1}(x)
                +5 H_{0,1,0,0,1}(x)
                +4 H_{0,1,0,1,1}(x)
                -8 H_{0,1,1,0,1}(x)
\nn\\&                
                +\big(
                        -24 H_{0,1,1}(x)
                        +23 \zeta_3
                        +H_{0,0,1}(x)
                \big) \zeta_2
                -3 H_{0,1}(x) \zeta_3
                -2 \zeta_5
        \big)
        +\frac{1}{8} \big(
                452
                +15 \ixo
                +94 x
\nn\\&                
                -2172 x^2
                -129 x^3
        \big) H_{0,0,0,1}(x)
        +\frac{1}{4} \big(
                852
                -5 \ixo
                +302 x
                -2700 x^2
                -189 x^3
        \big) H_{0,0,1,1}(x)
\nn\\&        
        +\frac{1}{4} \big(
                604
                +5 \ixo
                +930 x
                -636 x^2
                -33 x^3
        \big) H_{0,1,0,1}(x)
        +\frac{1}{16} \big(
                1020
                +17 \ixo
                +4342 x
\nn\\&                
                +4700 x^2
                +361 x^3
        \big) H_{0,1}(x) \zeta_2
        -\frac{27}{20} \big(
                32+163 x+224 x^2+16 x^3\big) \zeta_2^2
\big)
\nn\\&
+\iomx^3 \big(
        \frac{1}{96} \big(
                6429
                +1033 \ixo
                -29133 x
                -5833 x^2
        \big) H_{0,1}(x)
        +\frac{1}{4}
         \big(
                352
                +22 \ixo
                +171 x
\nn\\&                
                -155 x^2
        \big) H_{0,0,1}(x)
        +\frac{1}{6} \big(
                1632
                -50 \ixo
                -1875 x
                -805 x^2
        \big) H_{0,1,1}(x)
        +\frac{1}{12} \big(
                1275
                +115 \ixo
\nn\\&                
                -3900 x
                -928 x^2
        \big) H_{1,0,1}(x)
        +\frac{1}{2} \big(
                1089
                -43 \ixo
                +1815 x
                +163 x^2
        \big) H_{0,1,1,1}(x)
\nn\\&        
        +\frac{1}{4} \big(
                373
                +9 \ixo
                +837 x
                +41 x^2
        \big) H_{1,0,0,1}(x)
        +\frac{1}{2} \big(
                235
                -25 \ixo
                +249 x
                +45 x^2
        \big) H_{1,0,1,1}(x)
\nn\\&        
        +\frac{1}{2} \big(
                -287
                -11 \ixo
                -681 x
                -29 x^2
        \big) H_{1,1,0,1}(x)
        +\big(
                -88
                -\frac{1385 x}{2}
                -79 x^2
                +\big(
                        -\frac{441}{4}
\nn\\&                        
                        +\frac{39 \ixo}{4}
                        -688 x
                        -71 x^2
                \big) H_1(x)
                +\frac{1}{8} \big(
                        -3639
                        -67 \ixo
                        -7977 x
                        -413 x^2
                \big) H_{1,1}(x)
        \big) \zeta_2
\nn\\&        
        -\big(
                 26
                +141 x
                +22 x^2
                +\frac{1}{8} \big(
                         393
                        +29 \ixo
                        +1059 x
                        +31 x^2
                \big) H_1(x)
        \big) \zeta_3
\big)
-65 \ixo H_{1,1,1,1}(x)
 \Big\}
\nn\\&
+ \Big\{
\iomx \big(
        -
        \frac{583}{12}
        +\frac{1}{576} (-55235-30669 \ixo) H_1(x)
\big)
+\iomx^2 \big(
        \frac{1}{432} \big(
                -711962
                +47383 \ixo
\nn\\&                
                +2988 \ixtw
                -288377 x
        \big) H_{1,1}(x)
        +\frac{1}{72} \big(
                33346
                +14923 \ixo
                +360 \ixtw
                +7099 x
        \big) H_{1,1,1}(x)
\nn\\&        
        +700 (7+2 x) H_{1,1,1,1}(x)
\big)
+\iomx^3 \big(
        \frac{1}{864} \big(
                -305541
                -42065 \ixo
                +1548981 x
\nn\\&                
                +698561 x^2
        \big) H_{0,1}(x)
        +\frac{1}{288} \big(
                -230689
                -5085 \ixo
                -464655 x
                -40451 x^2
        \big) H_{0,0,1}(x)
\nn\\&        
        +\frac{1}{144} \big(
                -156013
                +13351 \ixo
                +288 \ixtw
                -513819 x
                -52287 x^2
        \big) H_{0,1,1}(x)
\nn\\&        
        +\frac{1}{144} \big(
                101903
                -1453 \ixo
                -72 \ixtw
                +161721 x
                +18269 x^2
        \big) H_{1,0,1}(x)
        +\frac{1}{3} \big(
                3395
                -19 \ixo
\nn\\&                
                -13653 x
                -3979 x^2
        \big) H_{0,1,1,1}(x)
        +\frac{1}{12} \big(
                -351
                +305 \ixo
                -8892 x
                -2852 x^2
        \big) H_{1,0,0,1}(x)
\nn\\&        
        +\big(
                \frac{5035}{3}
                +71 \ixo
                -1832 x
                -\frac{1948 x^2}{3}
        \big) H_{1,0,1,1}(x)
        +\frac{1}{6} \big(
                6040
                +364 \ixo
                -4857 x
\nn\\&                
                -2429 x^2
        \big) H_{1,1,0,1}(x)
        +\big(
                4495
                -165 \ixo
                +7605 x
                +665 x^2
        \big) H_{0,1,1,1,1}(x)
        +
        \frac{1}{4} \big(
                2133
                +57 \ixo
\nn\\&                
                +4277 x
                +97 x^2
        \big) H_{1,0,0,0,1}(x)
        +\frac{1}{2} \big(
                2535
                +3 \ixo
                +4601 x
                +181 x^2
        \big) H_{1,0,0,1,1}(x)
        +\frac{1}{2} \big(
                787
\nn\\&                
                +7 \ixo
                +1937 x
                +161 x^2
        \big) H_{1,0,1,0,1}(x)
        +\big(
                2043
                -121 \ixo
                +3039 x
                +331 x^2
        \big) H_{1,0,1,1,1}(x)
\nn\\&        
        +\frac{1}{2} \big(
                1045
                +25 \ixo
                +2343 x
                +115 x^2
        \big) H_{1,1,0,0,1}(x)
        +\big(
                397
                -79 \ixo
                +87 x
\nn\\&                
                +99 x^2
        \big) H_{1,1,0,1,1}(x)
        -13 \big(
                71
                +3 \ixo
                +171 x
                +7 x^2
        \big) H_{1,1,1,0,1}(x)
\nn\\&        
        +\big(
                \ixo
                +\frac{1}{36} \big(
                        39067+117378 x+43427 x^2\big)
                +\frac{1}{576} \big(
                        958821
                        -30967 \ixo
                        -576 \ixtw
\nn\\&                        
                        +2171547 x
                        +244855 x^2
                \big) H_1(x)
                +\frac{1}{6} \big(
                        4458
                        +349 \ixo
                        -555 x
                        -1066 x^2
                \big) H_{1,1}(x)
\nn\\&                
                +\frac{1}{8} \big(
                        -4417
                        -69 \ixo
                        -6213 x
                        +295 x^2
                \big) H_{1,0,1}(x)
                -\frac{3}{4} \big(
                        3923
                        +79 \ixo
                        +8661 x
\nn\\&                        
                        +441 x^2
                \big) H_{1,1,1}(x)
        \big) \zeta_2
        +\frac{1}{320} \big(
                -101245
                -897 \ixo
                -334035 x
                -41039 x^2
        \big) H_1(x) \zeta_2^2
\nn\\&        
        +\big(
                -\frac{461}{6}
                -180 x
                -\frac{323 x^2}
                {3}
                +
                \frac{1}{36} \big(
                        -3732
                        +16 \ixo
                        -30489 x
                        -6133 x^2
                \big) H_1(x)
\nn\\&                
                +\frac{1}{4} \big(
                        -1545
                        -125 \ixo
                        -4263 x
                        -115 x^2
                \big) H_{1,1}(x)
        \big) \zeta_3
\big)
\nn\\&
+\iomx^4 \big(
        \big(
                6+37 x+20 x^2
        \big)
\big(144 H_{0,0,0,1,1,1}(x)
                +200 H_{0,0,1,1,1,1}(x)
                +84 H_{0,1,0,1,1,1}(x)
\nn\\&                
                +28 H_{0,1,1,0,0,1}(x)
                +8 H_{0,1,1,0,1,1}(x)
                -52 H_{0,1,1,1,0,1}(x)
                +4 H_{1,0,0,0,0,1}(x)
\nn\\&                
                +16 H_{1,0,0,0,1,1}(x)
                -4 H_{1,0,0,1,0,1}(x)
                -4 H_{1,0,1,0,0,1}(x)
                +8 H_{1,0,1,0,1,1}(x)
                -4 H_{1,0,1,1,0,1}(x)
\nn\\&                
                +\big(
                        -156 H_{0,1,1,1}(x)
                        -12 H_{1,0,0,1}(x)
                        +4 H_{1,0,1,1}(x)
                        +8 H_1(x) \zeta_3
                \big) \zeta_2
                +\big(
                        -24 H_{0,1,1}(x)
\nn\\&                        
                        -12 H_{1,0,1}(x)
                \big) \zeta_3
                +22 H_1(x) \zeta_5
        \big)
        +\frac{1}{36} \big(
                -11772
                +159 \ixo
                -51220 x
                -7670 x^2
\nn\\&                
                +4191 x^3
        \big) H_{0,0,0,1}(x)
        +\frac{1}{18} \big(
                5304
                +1011 \ixo
                -76258 x
                -4778 x^2
                +8409 x^3
        \big) H_{0,0,1,1}(x)
\nn\\&        
        +\frac{1}{36} \big(
                -1002
                -645 \ixo
                +7879 x
                +49892 x^2
                +10188 x^3
        \big) H_{0,1,0,1}(x)
        +\frac{1}{24} \big(
                -708
                +81 \ixo
\nn\\&                
                -20510 x
                -22564 x^2
                -147 x^3
        \big) H_{0,0,0,0,1}
        (x)
        +
        \frac{1}{12} \big(
                10332
                +285 \ixo
                -5078 x
                -42004 x^2
\nn\\&                
                -903 x^3
        \big) H_{0,0,0,1,1}(x)
        +\big(
                15
                -\frac{15 \ixo}{4}
                -\frac{7349 x}{6}
                -\frac{6833 x^2}{3}
                -\frac{489 x^3}{4}
        \big) H_{0,0,1,0,1}(x)
\nn\\&        
        +\frac{1}{6} \big(
                11604
                -201 \ixo
                -14878 x
                -53564 x^2
                -3225 x^3
        \big) H_{0,0,1,1,1}(x)
        +\frac{1}{4} \big(
                572
                +37 \ixo
\nn\\&                
                -4134 x
                -8124 x^2
                -345 x^3
        \big) H_{0,1,0,0,1}(x)
        +\frac{1}{6} \big(
                9852
                -51 \ixo
                +7138 x
                -20260 x^2
\nn\\&                
                -1071 x^3
        \big) H_{0,1,0,1,1}(x)
        +\frac{1}{6} \big(
                5484
                -45 \ixo
                +13376 x
                -1376 x^2
                -177 x^3
        \big) H_{0,1,1,0,1}(x)
\nn\\&        
        +(3+4 x) (14+13 x) H_{0,0,0,0,0,1}(x)
        +2 (18+11 x) (11+52 x) H_{0,0,0,0,1,1}(x)
\nn\\&        
        +(1+8 x) (102+59 x) H_{0,0,0,1,0,1}(x)
        +\big(
                138+769 x+416 x^2\big) H_{0,0,1,0,0,1}(x)
\nn\\&                
        +8 \big(
                63+368 x+199 x^2\big) H_{0,0,1,0,1,1}(x)
        +24 \big(
                7+50 x+27 x^2\big) H_{0,0,1,1,0,1}(x)
\nn\\&                
        +\big(
                174+991 x+536 x^2\big) H_{0,1,0,0,0,1}(x)
        +4 \big(
                96+551 x+298 x^2\big) H_{0,1,0,0,1,1}(x)
\nn\\&                
        +6 \big(
                20+137 x+74 x^2\big) H_{0,1,0,1,0,1}(x)
        +\big(
                \frac{1}{24} \big(
                        -7086
                        -1267 \ixo
                        +80892 x
                        +56850 x^2
\nn\\&                        
                        +3235 x^3
                \big) H_{0,1}(x)
                +\frac{1}{48} \big(
                        -24636
                        -729 \ixo
                        -47266 x
                        -26156 x^2
                        -6549 x^3
                \big) H_{0,0,1}(x)
\nn\\&                
                +\frac{1}{8} \big(
                        4476
                        -79 \ixo
                        +41306 x
                        +37852 x^2
                        +1333 x^3
                \big) H_{0,1,1}(x)
\nn\\&                
                +\frac{3}{2} (6+x) (-13+4 x) H_{0,0,0,1}(x)
                +\big(
                        -174-581 x-316 x^2\big) H_{0,0,1,1}(x)
\nn\\&                        
                +\frac{1}{2} \big(
                        -354-1691 x-916 x^2\big) H_{0,1,0,1}(x)
                -\big(
                         748+\frac{22003 x}{6}+\frac{11428 x^2}{3}+192 x^3\big) \zeta_3
        \big) \zeta_2
\nn\\&        
        -\big(
                \frac{1}{20} \big(
                         17749+79906 x+10300 x^2-8487 x^3\big)
                +\frac{1}{10} \big(
                         942+8023 x+4328 x^2\big) H_{0,1}(x)
        \big) \zeta_2^2
\nn\\&        
        +\frac{1}{420} \big(
                209550+1224821 x+662332 x^2\big) \zeta_2^3
        +\big(
                \frac{1}{24} \big(
                        1788
                        -159 \ixo
                        -2194 x
                        -7604 x^2
\nn\\&                        
                        -579 x^3
                \big) H_{0,1}(x)
                +\big(
                        18+193 x+104 x^2\big) H_{0,0,1}(x)
        \big) \zeta_3
        +\frac{1}{2} \big(
                -18-193 x-104 x^2\big) \zeta_3^2
\nn\\&                
        +\big(
                456+\frac{13331 x}{6}+\frac{5993 x^2}{3}+114 x^3\big) \zeta_5
\big)
-422 \ixo H_{1,1,1,1,1}(x)
 \Big\}
\bigg] \,.
\end{align}
% 
% 
% 
% % % % % % % % % % % % % % % % % % % % % % % % % % % % 
% 
The form factor $G_3^{(3)}$ in the color-planar limit, is given by

% 
% 
% % % % % % % % % % % % % % % % % % % % 
% 
The form factor $S^{(3)}$ in the color-planar limit, is given by
\begin{align}
S^{(3)} &= 
 N_C^3 \bigg[
 \frac{1}{\ep^6} \Big\{
 -\frac{1}{48}
 \Big\}
+  \frac{1}{\ep^5} \Big\{
-\frac{27}{32}-\frac{1}{8} H_1(x)
 \Big\}
+  \frac{1}{\ep^4} \Big\{  
-\frac{37055}{5184}
+\frac{1}{24} (-70-3 \ixo) H_1(x)
\nn\\&
-\frac{1}{8} H_{0,1}(x)
-\frac{3}{4} H_{1,1}(x)
-\frac{5}{32} \zeta_2
 \Big\}
+  \frac{1}{\ep^3} \Big\{  
-\frac{123775}{31104}
+\frac{1}{864} (-7067-1836 \ixo) H_1(x)
\nn\\&
+\frac{1}{8} (-16-\ixo) H_{0,1}(x)
+\frac{1}{12} (-107-15 \ixo) H_{1,1}(x)
-\frac{1}{8} H_{0,0,1}(x)
-\frac{5}{4} H_{0,1,1}(x)
\nn\\&
-\frac{3}{4} H_{1,0,1}(x)
-\frac{9}{2} H_{1,1,1}(x)
-\frac{29}{48} \zeta_3
+\frac{\zeta_2 \big(
        -3191-1620 H_1(x)\big)}{1728}
 \Big\}
\nn\\&
+  \frac{1}{\ep^2} \Big\{  
\frac{2746}{243}
+\iomx \big(
        \frac{1}{288} (-73
        -228 \ixo
        -419 x
        ) H_{0,1}(x)
        +\frac{1}{24} (11
        +9 \ixo
        +28 x
        ) H_{0,0,1}(x)
\nn\\&        
        +\big(
                \frac{1}{4}
                -\frac{9 \ixo}{4}
                +6 x
        \big) H_{0,1,1}(x)
        +\frac{1}{12} (-74
        -9 \ixo
        +59 x
        ) H_{1,0,1}(x)
\nn\\&        
        +\big(
                \frac{1}{10368} (-38981-18043 x)
                +\frac{1}{144} (-1303
                -135 \ixo
                +574 x
                ) H_1(x)
        \big) \zeta_2
\big)
\nn\\&
+\iomx^2 \big(
        \frac{1}{8} \big(
                15-6 x+7 x^2\big) H_{0,0,0,1}(x)
        +\frac{1}{4} \big(
                3+18 x-5 x^2\big) H_{0,0,1,1}(x)
\nn\\&                
        +\frac{1}{4} \big(
                -9+6 x-5 x^2\big) H_{0,1,0,1}(x)
        -\frac{3}{16} \big(
                29-10 x+13 x^2\big) H_{0,1}(x) \zeta_2
\nn\\&                
        +\frac{1}{5760} \big(
                15229+16198 x-323 x^2\big) \zeta_2^2
\big)
+\frac{1}{2592} (33781+3015 \ixo) H_1(x)
\nn\\&
+\frac{1}{48} \big(
        305-300 \ixo-48 \ixtw\big) H_{1,1}(x)
+\frac{1}{6} (-115-57 \ixo) H_{1,1,1}(x)
-\frac{19}{2} H_{0,1,1,1}(x)
\nn\\&
-\frac{3}{4} H_{1,0,0,1}(x)
-\frac{15}{2} H_{1,0,1,1}(x)
-\frac{9}{2} H_{1,1,0,1}(x)
-27 H_{1,1,1,1}(x)
-\frac{45}{8} H_{1,1}(x) \zeta_2
\nn\\&
+\frac{1}{864} \zeta_3 \big(
        -4993-3132 H_1(x)\big)
 \Big\}
+  \frac{1}{\ep} \Big\{  
\frac{1761547}{69984}
+\iomx \big(
        \frac{1}{864} (-1802
        +7677 \ixo
\nn\\&        
        -14299 x
        ) H_{0,1}(x)
        +\frac{1}{96} (349
        +300 \ixo
        -697 x
        ) H_{0,0,1}(x)
        +\frac{1}{48} \big(
                1449
                +92 \ixo
                -96 \ixtw
\nn\\&                
                -2213 x
        \big) H_{0,1,1}(x)
        +\frac{1}{48} \big(
                413
                +448 \ixo
                -48 \ixtw
                -1149 x
        \big) H_{1,0,1}(x)
        +\frac{1}{2} (171
        -43 \ixo
\nn\\&        
        -32 x
        ) H_{0,1,1,1}(x)
        +\frac{1}{12} (94
        +27 \ixo
        -x
        ) H_{1,0,0,1}(x)
        +\frac{1}{2} (48
        -25 \ixo
        -7 x
        ) H_{1,0,1,1}(x)
\nn\\&        
        +\frac{1}{6} (-52
        -33 \ixo
        -11 x
        ) H_{1,1,0,1}(x)
        +\big(
                \frac{1}{62208} (-120005-766459 x)
\nn\\&                
                +\frac{1}{1728} (-9677
                +11340 \ixo
                -31039 x
                ) H_1(x)
                +\frac{1}{24} (-868
                -201 \ixo
                -83 x
                ) H_{1,1}(x)
        \big) \zeta_2
\nn\\&        
        +\big(
                \frac{1}{1728} (5753-11801 x)
                +\frac{1}{8} (-93
                -29 \ixo
                +74 x
                ) H_1(x)
        \big) \zeta_3
\big)
+\iomx^2 \big(
        \frac{1}{8} \big(
                22
                +15 \ixo
\nn\\&                
                -149 x
                +32 x^2
        \big) H_{0,0,0,1}(x)
        +\frac{1}{12} \big(
                218
                -15 \ixo
                -475 x
                +32 x^2
        \big) H_{0,0,1,1}(x)
\nn\\&        
        +\frac{1}{12} \big(
                232
                +15 \ixo
                -155 x
                +28 x^2
        \big) H_{0,1,0,1}
        (x)
        +
        \frac{1}{8} \big(
                63-30 x+31 x^2\big) H_{0,0,0,0,1}(x)
\nn\\&                
        +\frac{1}{4} \big(
                67+10 x+19 x^2\big) H_{0,0,0,1,1}(x)
        +\frac{1}{4} \big(
                47-10 x+19 x^2\big) H_{0,0,1,0,1}(x)
\nn\\&                
        +\frac{1}{2} \big(
                29+86 x-19 x^2\big) H_{0,0,1,1,1}(x)
        +\frac{1}{4} \big(
                39-18 x+19 x^2\big) H_{0,1,0,0,1}(x)
\nn\\&                
        +\frac{1}{2} \big(
                -13+50 x-21 x^2\big) H_{0,1,0,1,1}(x)
        +\frac{1}{2} \big(
                -35+22 x-19 x^2\big) H_{0,1,1,0,1}(x)
\nn\\&                
        +\big(
                \frac{1}{16} \big(
                        218
                        +17 \ixo
                        +289 x
                        -44 x^2
                \big) H_{0,1}(x)
                +\frac{1}{16} \big(
                        41-34 x+25 x^2\big) H_{0,0,1}(x)
\nn\\&                        
                +\frac{1}{8} \big(
                        -355+134 x-163 x^2\big) H_{0,1,1}(x)
                +\frac{1}{288} \big(
                        9823+226 x+3199 x^2\big) \zeta_3
        \big) \zeta_2
\nn\\&        
        +\frac{1}{11520} \big(
                -178171-125770 x-7099 x^2\big) \zeta_2^2
        +\frac{1}{8} \big(
                -65+58 x-41 x^2\big) H_{0,1}(x) \zeta_3
\nn\\&                
        +\frac{1}{240} \big(
                -5087+8734 x-4607 x^2\big) \zeta_5
\big)
+\frac{1}{1728} (-3309+59072 \ixo) H_1(x)
\nn\\&
+\frac{1}{48} \big(
        1243+2545 \ixo-216 \ixtw\big) H_{1,1}(x)
+\frac{1}{24} \big(
        2233+444 \ixo-288 \ixtw\big) H_{1,1,1}(x)
\nn\\&        
+\frac{1}{3} (67-195 \ixo) H_{1,1,1,1}(x)
-65 H_{0,1,1,1,1}(x)
+\frac{9}{4} H_{1,0,0,0,1}(x)
-\frac{27}{2} H_{1,0,0,1,1}(x)
\nn\\&
-\frac{9}{2} H_{1,0,1,0,1}(x)
-57 H_{1,0,1,1,1}(x)
-\frac{9}{2} H_{1,1,0,0,1}(x)
-45 H_{1,1,0,1,1}(x)
-27 H_{1,1,1,0,1}(x)
\nn\\&
-162 H_{1,1,1,1,1}(x)
-\big(
         \frac{45}{8} H_{1,0,1}(x)
        +\frac{135}{4} H_{1,1,1}(x)
\big) \zeta_2
-\frac{8099}{960} H_1(x) \zeta_2^2
-\frac{87}{4} H_{1,1}(x) \zeta_3
 \Big\}
\nn\\&
+ \Big\{
\frac{4233773}{419904}
+\iomx \big(
        \frac{1}{1728} (120419
        -24352 \ixo
        +105149 x
        ) H_{0,1}(x)
        +\frac{1}{864} (-85898
\nn\\&        
        -14067 \ixo
        +5957 x
        ) H_{0,0,1}(x)
        +\frac{1}{432} \big(
                -143750
                +26553 \ixo
                +288 \ixtw
                +33989 x
        \big) H_{0,1,1}(x)
\nn\\&        
        +\frac{1}{432} \big(
                33010
                +14541 \ixo
                -2880 \ixtw
                +3521 x
        \big) H_{1,0,1}(x)
        +\big(
                \frac{985 \ixo}{6}
                -32 \ixtw
\nn\\&                
                +\frac{1}{72} (-14441-16915 x)
        \big) H_{0,1,1,1}(x)
        +\frac{1}{144} (-4105
        +2328 \ixo
        -4559 x
        ) H_{1,0,0,1}(x)
\nn\\&        
        +\frac{1}{72} \big(
                2471
                +9756 \ixo
                -1296 \ixtw
                -13715 x
        \big) H_{1,0,1,1}(x)
        +\frac{1}{72} \big(
                4655
                +6132 \ixo
                -720 \ixtw
\nn\\&                
                -8183 x
        \big) H_{1,1,0,1}(x)
        -5 (-185
        +33 \ixo
        +72 x
        ) H_{0,1,1,1,1}(x)
\nn\\&        
        +\big(
                63
                +\frac{69 \ixo}{4}
                -\frac{49 x}{4}
        \big) H_{1,0,0,0,1}(x)
        +\frac{1}{2} (418
        +15 \ixo
        -137 x
        ) H_{1,0,0,1,1}(x)
\nn\\&        
        +\frac{1}{2} (151
        +\ixo
        -92 x
        ) H_{1,0,1,0,1}(x)
        +\frac{1}{3} (1666
        -363 \ixo
        -799 x
        ) H_{1,0,1,1,1}(x)
\nn\\&        
        +\frac{1}{6} (452
        +75 \ixo
        -191 x
        ) H_{1,1,0,0,1}
        (x)
        +(298
        -79 \ixo
        -203 x
        ) H_{1,1,0,1,1}(x)
\nn\\&        
        +\big(
                \frac{232}{3}
                -39 \ixo
                -\frac{427 x}{3}
        \big) H_{1,1,1,0,1}(x)
        +\big(
                \frac{1}{23328} (5042833
                +124416 \ixo
                +2370287 x
                )
\nn\\&                
                +\frac{1}{5184} \big(
                        1696418
                        +10629 \ixo
                        -17280 \ixtw
                        +10585 x
                \big) H_1(x)
                +\frac{1}{96} \big(
                        10951
                        +4684 \ixo
\nn\\&                        
                        -336 \ixtw
                        -6035 x
                \big) H_{1,1}(x)
                +\frac{1}{8} (-342
                -141 \ixo
                -229 x
                ) H_{1,0,1}(x)
\nn\\&                
                -\frac{3}{4} (166
                +79 \ixo
                +171 x
                ) H_{1,1,1}(x)
        \big) \zeta_2
        +\frac{1}{320} (6439
        +1695 \ixo
        -9542 x
        ) H_1(x) \zeta_2^2
\nn\\&        
        +\big(
                \frac{1}{31104} (1417255
                +62208 \ixo
                -2337415 x
                )
                +\frac{1}{864} \big(
                        21247
                        +14532 \ixo
                        -1728 \ixtw
\nn\\&                        
                        -51619 x
                \big) H_1(x)
                +\frac{1}{36} (664
                -1125 \ixo
                -1267 x
                ) H_{1,1}(x)
        \big) \zeta_3
\big)
\nn\\&
+\iomx^2 \big(
        \frac{1}{288} \big(
                -25797
                +156 \ixo
                +7686 x
                -4861 x^2
        \big) H_{0,0,0,1}(x)
        +\frac{1}{144} \big(
                -27593
\nn\\&                
                +5964 \ixo
                -11474 x
                +10287 x^2
        \big) H_{0,0,1,1}(x)
        +\frac{1}{144} \big(
                -865
                +912 \ixo
                -432 \ixtw
                +4106 x
\nn\\&                
                +7687 x^2
        \big) H_{0,1,0,1}(x)
        +\frac{1}{24} \big(
                -602
                +81 \ixo
                -935 x
                +12 x^2
        \big) H_{0,0,0,0,1}(x)
\nn\\&        
        +
        \frac{1}{12} \big(
                110
                +357 \ixo
                -2347 x
                +124 x^2
        \big) H_{0,0,0,1,1}(x)
        +\frac{1}{12} \big(
                -440
                -9 \ixo
                -1475 x
\nn\\&                
                +152 x^2
        \big) H_{0,0,1,0,1}(x)
        +\frac{1}{6} \big(
                1054
                -93 \ixo
                -4145 x
                +596 x^2
        \big) H_{0,0,1,1,1}(x)
\nn\\&        
        +\frac{1}{12} \big(
                -352
                +147 \ixo
                -1123 x
                +56 x^2
        \big) H_{0,1,0,0,1}(x)
        +\frac{1}{6} \big(
                1560
                -51 \ixo
                -2223 x
\nn\\&                  
                +518 x^2
        \big) H_{0,1,0,1,1}(x)
        +\frac{1}{6} \big(
                1357
                -81 \ixo
                -953 x
                +422 x^2
        \big) H_{0,1,1,0,1}(x)
\nn\\&        
        +\frac{1}{8} \big(
                111-54 x+23 x^2\big) H_{0,0,0,0,0,1}(x)
        +\frac{1}{4} \big(
                467-142 x+171 x^2\big) H_{0,0,0,0,1,1}(x)
\nn\\&                
        +\frac{1}{4} \big(
                111-18 x+67 x^2\big) H_{0,0,0,1,0,1}(x)
        +\frac{1}{2} \big(
                341+182 x+53 x^2\big) H_{0,0,0,1,1,1}(x)
\nn\\&                
        +\frac{1}{4} \big(
                175-74 x+75 x^2\big) H_{0,0,1,0,0,1}(x)
        +\frac{1}{2} \big(
                235+34 x+59 x^2\big) H_{0,0,1,0,1,1}(x)
\nn\\&                
        +\frac{3}{2} \big(
                31-6 x+15 x^2\big) H_{0,0,1,1,0,1}(x)
        -5 \big(
                -27-66 x+13 x^2\big) H_{0,0,1,1,1,1}(x)
\nn\\&                
        +\frac{1}{4} \big(
                231-114 x+107 x^2\big) H_{0,1,0,0,0,1}(x)
        +\frac{1}{2} \big(
                195-6 x+59 x^2\big) H_{0,1,0,0,1,1}(x)
\nn\\&                
        +
        \frac{1}{2} \big(
                67-14 x+31 x^2\big) H_{0,1,0,1,0,1}(x)
        +\big(
                5+242 x-79 x^2\big) H_{0,1,0,1,1,1}(x)
\nn\\&                
        +\frac{1}{2} \big(
                109-50 x+53 x^2\big) H_{0,1,1,0,0,1}(x)
        +\big(
                -67+158 x-75 x^2\big) H_{0,1,1,0,1,1}(x)
\nn\\&                
        -13 \big(
                9-6 x+5 x^2\big) H_{0,1,1,1,0,1}(x)
        +\frac{1}{4} \big(
                77-106 x+61 x^2\big) H_{1,0,0,0,0,1}(x)
\nn\\&                
        +\frac{1}{2} \big(
                49-2 x+17 x^2\big) H_{1,0,0,0,1,1}(x)
        +\frac{1}{2} \big(
                -1-22 x+7 x^2\big) H_{1,0,0,1,0,1}(x)
\nn\\&                
        +\frac{1}{2} \big(
                11-46 x+19 x^2\big) H_{1,0,1,0,0,1}(x)
        +\big(
                -59+142 x-67 x^2\big) H_{1,0,1,0,1,1}(x)
\nn\\&                
        +\big(
                -41+70 x-37 x^2\big) H_{1,0,1,1,0,1}(x)
        +\big(
                \frac{1}{576} \big(
                        76337
                        -15324 \ixo
                        -1152 \ixtw
                        +50018 x
\nn\\&                        
                        +27017 x^2
                \big) H_{0,1}(x)
                +\frac{1}{48} \big(
                        -3806
                        -729 \ixo
                        -641 x
                        +344 x^2
                \big) H_{0,0,1}(x)
\nn\\&                
                +\frac{1}{24} \big(
                        6754
                        -669 \ixo
                        +2155 x
                        +1708 x^2
                \big) H_{0,1,1}(x)
                -\frac{3}{16} \big(
                        205-98 x+5 x^2\big) H_{0,0,0,1}(x)
\nn\\&                        
                +\frac{1}{8} \big(
                        -331-34 x-3 x^2\big) H_{0,0,1,1}(x)
                +\frac{1}{8} \big(
                        -423+138 x-139 x^2\big) H_{0,1,0,1}(x)
\nn\\&                        
                -\frac{3}{4} \big(
                        391-158 x+183 x^2\big) H_{0,1,1,1}(x)
                -
                \frac{3}{8} \big(
                        47+2 x+15 x^2\big) H_{1,0,0,1}(x)
\nn\\&                        
                +\frac{1}{4} \big(
                        -169+386 x-185 x^2\big) H_{1,0,1,1}(x)
                +\big(
                        \frac{1}{192} \big(
                                -38247-22866 x+857 x^2\big)
\nn\\&                                
                        +\frac{1}{48} \big(
                                655-158 x+271 x^2\big) H_1(x)
                \big) \zeta_3
        \big) \zeta_2
        +\big(
                \frac{1}{69120} (-11522237-2723462 x
\nn\\&                
                -539069 x^2)             
                +\frac{1}{320} \big(
                        -8433+1794 x-5137 x^2\big) H_{0,1}(x)
        \big) \zeta_2^2
        +\frac{1}{241920} \big(
                25868957
\nn\\&                
                +8612486 x
                +4805213 x^2\big) \zeta_2^3
        +\big(
                \frac{1}{72} \big(
                        2480
                        -477 \ixo
                        -2413 x
                        +386 x^2
                \big) H_{0,1}(x)
\nn\\&                
                +\frac{1}{8} \big(
                        31+10 x+23 x^2\big) H_{0,0,1}(x)
                +\frac{1}{4} \big(
                        -269+250 x-173 x^2\big) H_{0,1,1}(x)
\nn\\&                        
                -\frac{3}{4} \big(
                        61-74 x+45 x^2\big) H_{1,0,1}(x)
        \big) \zeta_3
        +\frac{1}{288} \big(
                -4759+8222 x-4615 x^2\big) \zeta_3^2
\nn\\&                
        +\big(
                \frac{1}{96} \big(
                        8241+7566 x-2183 x^2\big)
                -\frac{11}{40} \big(
                        237-714 x+317 x^2\big) H_1(x)
        \big) \zeta_5
\big)
\nn\\&
+\frac{1}{93312} (5454773-6967134 \ixo) H_1(x)
-\frac{1}{2592} \big(
         227743+90336 \ixo-84744 \ixtw\big) H_{1,1}(x)
\nn\\&         
+\big(
        -\frac{20345}{216}+\frac{18391 \ixo
        }{72}+\frac{56 \ixtw}{3}-6 \ixthr\big) H_{1,1,1}(x)
-\frac{5}{36} \big(
        -3199-2844 \ixo
\nn\\&        
        +720 \ixtw\big) H_{1,1,1,1}(x)
-\frac{2}{3} (-1025+633 \ixo) H_{1,1,1,1,1}(x)
-422 H_{0,1,1,1,1,1}(x)
\nn\\&
-129 H_{1,0,0,1,1,1}(x)
-390 H_{1,0,1,1,1,1}(x)
+\frac{27}{2} H_{1,1,0,0,0,1}(x)
-81 H_{1,1,0,0,1,1}(x)
\nn\\&
-27 H_{1,1,0,1,0,1}(x)
-342 H_{1,1,0,1,1,1}(x)
-27 H_{1,1,1,0,0,1}(x)
-270 H_{1,1,1,0,1,1}(x)
\nn\\&
-162 H_{1,1,1,1,0,1}(x)
-972 H_{1,1,1,1,1,1}(x)
+\big(
        -\frac{135}{4} H_{1,1,0,1}(x)
        -\frac{405}{2} H_{1,1,1,1}(x)
\big) \zeta_2
\nn\\&
-\frac{8099}{160} H_{1,1}(x) \zeta_2^2
-\frac{261}{2} H_{1,1,1}(x) \zeta_3
 \Big\}
\bigg] \,.
\end{align}
% 
% 
% 
% 
% 
% 

% \newpage
\section{Four-loop form factors in the asymptotic limit}
\label{sec:res4L}
In this section, we present partial results for the four-loop form factors in the asymptotic limit ($\tilde{G}_1^{(4)}$ and $\tilde{S}^{(4)}$)\,.
With all the available components, we have obtained up to $\frac{1}{\ep^3}$ poles completely,
while partial results have been computed for $\frac{1}{\ep^2}$, $\frac{1}{\ep}$ and $\ep^0$ coefficients.
The light fermionic contributions are completely known for the $\frac{1}{\ep^2}$ pole, while 
$\mathcal{O}(L^0)$ and $\mathcal{O}(L^0,L^1)$ terms can not be predicted for the $\frac{1}{\ep}$ and ${\ep^0}$ coefficients,
respectively.
For the non-fermionic contributions, 
all the higher powers of $L$ can be predicted except for $\mathcal{O}(L^0)$ in $\frac{1}{\ep^2}$, 
$\mathcal{O}(L^0)$ and $\mathcal{O}(L^1)$ in $\frac{1}{\ep}$, 
$\mathcal{O}(L^0)$, $\mathcal{O}(L^1)$ and $\mathcal{O}(L^2)$ in ${\ep^0}$.
In the following, we mention the Casimirs appearing in the four-loop results
\begin{align}
 &
 N_A = (N_C^2-1)\,, \,
 N_F = N_C\,, \,
 \nn\\&
 \frac{d^{abcd}_F d^{abcd}_F}{N_F} = \frac{N_A (N_C^4-6 N_C^2+18)}{96 N_C^3}\,, \,
 \frac{d^{abcd}_A d^{abcd}_F}{N_F} = \frac{N_A (N_C^2+6)}{48} \,.
\end{align}
The result for $\tilde{G}_1^{(4)}$ is given below. 

% 
% 
% % % % % % % % % % % % % % % % % % % % % % % % % % % % % % % % % % % % 
% % % % % % % % % % % % % % % % % % % % % % % % % % % % % % % % % % % % 
% 
The result for $\tilde{S}^{(4)}$ is given below. 
%

\bibliography{main}
\bibliographystyle{JHEP}

\end{document}